\documentclass[prx,aps,epsfig,twocolumn,superscriptaddress]{revtex4} 

\usepackage{amsfonts}
\usepackage[dvips]{graphicx}
\usepackage{mathrsfs}
\usepackage{amsthm,amssymb}
\theoremstyle{plain}
\usepackage[intlimits]{amsmath}
\usepackage[colorlinks, citecolor=blue, linkcolor=blue, linktocpage]{hyperref}
\usepackage{graphicx}  
\usepackage{dcolumn}   
\usepackage{bm}        
\usepackage{amssymb}   
\usepackage{floatrow}
\usepackage{mathtools}
\usepackage{braket}
\usepackage[caption=false]{subfig} 
\usepackage{algorithm,algpseudocode}
\usepackage{boxedminipage}
\usepackage{varwidth} 
\usepackage{chngcntr} 

\usepackage[english]{babel}
\usepackage{blindtext}
\usepackage{makecell}

\usepackage{makecell}
\usepackage{array}
\newcolumntype{?}{!{\vrule width 2\arrayrulewidth}}

\usepackage{enumerate}
\usepackage{enumitem}

\makeatletter 
\renewcommand{\fnum@figure}{\textbf{Figure~\thefigure}}
\makeatother

\makeatletter 
\renewcommand{\fnum@algorithm}{\textbf{Algorithm~\thealgorithm}}
\makeatother

\makeatletter
\renewcommand{\thetable}{\arabic{table}} 
\renewcommand{\fnum@table}{\textbf{Table~\thetable}}
\makeatother

\newtheorem*{theorem*}{Theorem}

\theoremstyle{definition}

\newtheorem*{definition*}{Definition}

\makeatletter
\def\blfootnote{\xdef\@thefnmark{}\@footnotetext}
\makeatother

\newcommand{\one}{\mathbf{1}}

\newcommand{\CZ}{\textrm{CZ}}
\newcommand{\CCZ}{\textrm{CCZ}}

\begin{document}

\title{Hardware-Efficient, Fault-Tolerant Quantum Computation with Rydberg Atoms}

\author{Iris Cong}
\affiliation{Department of Physics, Harvard University, Cambridge, Massachusetts 02138, USA}

\author{Harry Levine}
\altaffiliation[Present address:]{AWS Center for Quantum Computing, Pasadena, CA, USA}
\affiliation{Department of Physics, Harvard University, Cambridge, Massachusetts 02138, USA}

\author{Alexander Keesling}
\affiliation{Department of Physics, Harvard University, Cambridge, Massachusetts 02138, USA}
\affiliation{QuEra Computing Inc., 1284 Soldiers Field Road, Boston, MA, 02135, USA}

\author{Dolev Bluvstein}
\affiliation{Department of Physics, Harvard University, Cambridge, Massachusetts 02138, USA}

\author{Sheng-Tao Wang}
\affiliation{QuEra Computing Inc., 1284 Soldiers Field Road, Boston, MA, 02135, USA}

\author{Mikhail D. Lukin}
\affiliation{Department of Physics, Harvard University, Cambridge, Massachusetts 02138, USA}

\date{\today}

\begin{abstract}
Neutral atom arrays have recently emerged as a promising platform for quantum information processing. One important remaining roadblock for the large-scale application of these systems is the ability to perform error-corrected quantum operations. To entangle the qubits in these systems, atoms are typically excited to Rydberg states, which could decay or give rise to various correlated errors that cannot be addressed directly through traditional methods of fault-tolerant quantum computation. In this work, we provide the first complete characterization of these sources of error in a neutral-atom quantum computer and propose hardware-efficient, fault-tolerant quantum computation schemes that mitigate them. Notably, we develop a novel and distinctly efficient method to address the most important errors associated with the decay of atomic qubits to states outside of the computational subspace. These  advances allow us to significantly reduce the resource cost for fault-tolerant quantum computation compared to existing, general-purpose schemes.
Our protocols can be implemented in the near-term using state-of-the-art neutral atom platforms with qubits encoded in both alkali and alkaline-earth 
atoms. 
\end{abstract}

\maketitle

\section{Introduction}

Neutral atom systems have recently emerged as a promising platform for quantum information processing. While the exceptional coherence times of their ground states enable long-lived quantum memories, fast, high-fidelity quantum operations can be achieved by individually addressing atoms with laser pulses and coupling them to highly-excited Rydberg states \cite{Jaksch00,Lukin01,Saffman10}. Furthermore, large numbers of individual neutral atoms can be deterministically arranged with arbitrary geometry in two- and three-dimensional systems~\cite{Barredo16,Ebadi20,Barredo18,Barredo20}. Recent experiments have demonstrated  quantum manipulation in large arrays of atoms for applications ranging from  quantum computing to  quantum simulations and quantum metrology~\cite{Bernien17,Levine18,Labuhn16,Maller15,Levine19,Graham19,Kim18,Keesling19,Omran19,Semeghini21,Bluvstein21b}. Several latest advances allowing for the dynamic reconfiguration of atoms have even led to realization of logical qubits encoded in color, surface, or toric codes, which is a first step to performing quantum error correction (QEC) on neutral atom platforms~\cite{Bluvstein21b}.

\begin{figure*}[t]
\includegraphics[width=\textwidth]{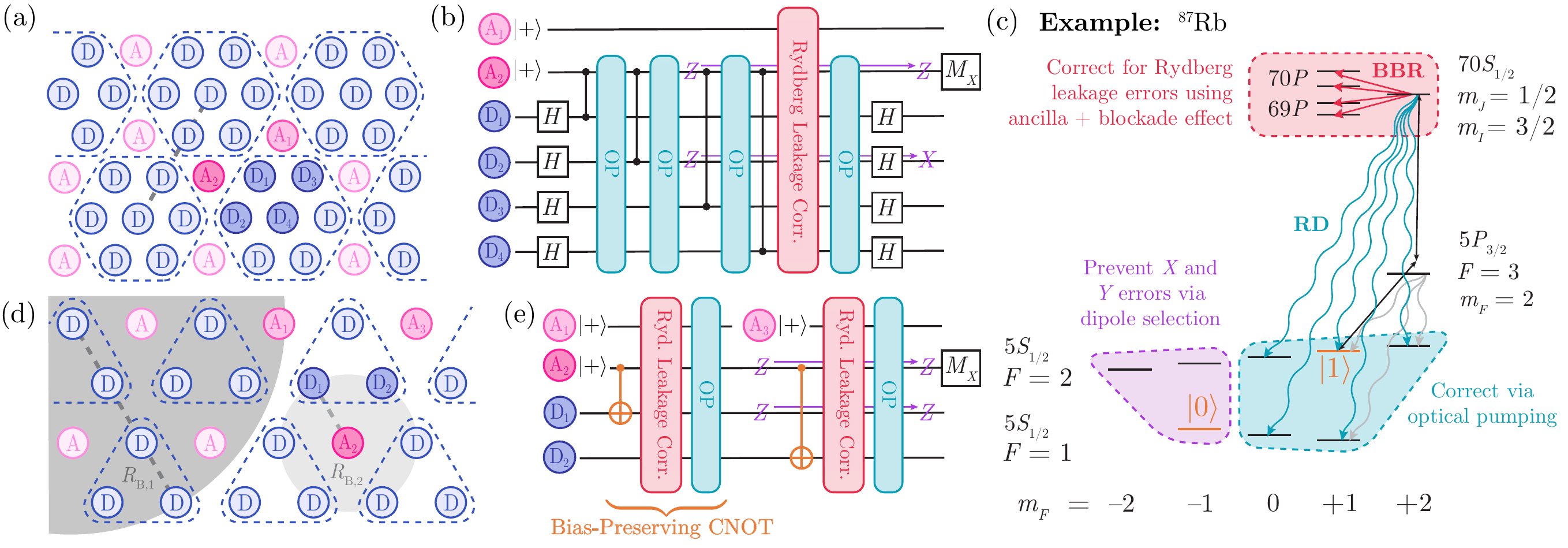}
\caption{Architecture for FTQC with Rydberg atoms. (a) Geometrical layout of atoms for FTQC using the seven-qubit encoding. Data (D, blue) and ancilla (A, pink) atoms are placed on the vertices of a triangular lattice, with seven data atoms comprising a logical qubit (blue dotted hexagons). The dotted grey line indicates the Rydberg interaction range required. (b) Circuit illustrating our procedure to measure a stabilizer operator, $X_1X_2X_3X_4$, for the seven-qubit code supported on the four data atoms highlighted in (a). Optical pumping (light blue, OP) is performed following every controlled-phase gate (black) to correct for leakage into other ground states. Ancilla qubit $A_2$ (darker pink) measures the stabilizer eigenvalue, while ancilla qubit $A_1$ (lighter pink) is used to detect and correct for Rydberg leakage errors (red). In this way, all gate errors are converted to Pauli-$Z$ type errors (purple) and do not spread to other qubits. (c) Level diagram showing an example encoding of a qubit in the hyperfine clock states of $^{87}$Rb. The dominant intrinsic errors for this encoding arise from blackbody radiation (BBR, red), radiative decay (RD, light blue), and intermediate state scattering (grey). Their effects can be determined via dipole selection rules (purple), and the relevant leakage errors can be corrected by making use of the Rydberg blockade effect or optical pumping.  (d) Geometrical layout for quantum computation with leading-order fault-tolerance using the three-atom encoding. Data and ancilla atoms are placed on the vertices of a triangular lattice, with three data atoms comprising a logical qubit (blue dotted triangles). In this case, two Rydberg states with different blockade radii,  $R_{B,1}$ and $R_{B,2}$ (dark and light grey, respectively) are required. (e) Our circuit for measuring a stabilizer operator, $X_1X_2$, of the repetition code supported on the two data atoms highlighted in (d). By combining a novel entangling pulse sequence with Rydberg leakage correction and optical pumping, we implement a bias-preserving CNOT gate (see Figure \ref{fig:bias-preserving-cnot-target}), allowing us to perform QEC without introducing $X$ or $Y$ errors at any point in the computation.}
\label{fig:overview-figure}
\end{figure*}

While current experiments are already demonstrating a remarkable level of quantum control, experimental imperfections such as Rydberg state decay will eventually limit the depth of accessible quantum operations. To scale up the computation size, it is therefore essential to consider QEC protocols~\cite{Nielsen00}. In particular, such protocols should be fault-tolerant and protect against the key sources of errors occurring within any of the computation, error detection, and encoding and decoding stages. Multiple fault-tolerant protocols have been proposed for generic quantum platforms~\cite{Shor96, Kitaev97, Gottesman10, Yoder17, Chao18, Chao18b, Reichardt18, Chamberland18, Chao20}, but they do not address certain  errors present in Rydberg atom setups. Indeed, Rydberg-atom QEC seems to face a daunting challenge at first glance: Rydberg states could decay into multiple other states, which not only results in leakage errors out of the computational space, but could also give rise to high-weight correlated errors from ensuing undesired blockade effects. 
Motivated by these considerations,  we  investigate the effects of these intrinsic errors. Remarkably, by utilizing the unique capabilities of Rydberg systems and the structure of the error model, we can design hardware-efficient, fault-tolerant quantum computation (FTQC) schemes that address these errors despite the aforementioned challenges (Figure~\ref{fig:overview-figure}). This tailored FTQC approach can even be much more resource efficient than generic proposals~\cite{Michael16,Chamberland20} (Tables~\ref{tab:stab-comparison} and \ref{tab:gate-comparison}), which often require a larger number of qubits and quantum operations with smaller threshold error than what is achievable in near-term experiments to perform non-Clifford logical operations, either directly~\cite{Yoder16,Chao18b} or by using state distillation~\cite{Shor96,Bravyi05}. The high overhead associated with such protocols is why experimental demonstrations of QEC have thus far been limited to only one or two logical qubits~\cite{Nigg14,Rosenblum18,Campagne20,DeNeeve20,Egan21}.

In this work, we first provide a detailed understanding, from the QEC perspective, of the errors arising from the finite lifetime of the Rydberg state or from imperfections in Rydberg laser pulses. 
We then show that nine atoms---seven data qubits and two ancilla qubits---are sufficient to encode each logical qubit fault-tolerantly based on the seven-qubit Steane code \cite{Steane96}; we demonstrate how a universal set of fault-tolerant quantum operations can be performed.
For atomic species with sufficiently large nuclear spin and high-fidelity ground-state operations, we show that quantum computation with leading-order fault-tolerance can be achieved even using a simple three-atom repetition code  \footnote{The three-qubit repetition code cannot correct any Pauli-$X$ errors, so it cannot be used for FTQC in typical setups. However, because the error model for Rydberg-atom setups does not contain any Pauli-$X$ errors at the leading order (as shown in Section~\ref{sec:error-models}), the repetition code is applicable in these platforms. We thus use the term ``leading-order fault-tolerance'' when describing our Ryd-3 protocol to emphasize this point explicitly.}. We find that both the seven-atom and three-atom codes can be implemented on scalable geometries with atoms placed in a triangular lattice configuration (Figure~\ref{fig:overview-figure}a,d), allowing for their demonstration and study in  near-term experiments. 

Our work provides an important advance over prior methods
by introducing a novel and distinctly efficient approach to address the leakage of qubits out of the computational subspace. 
For traditional QEC proposals, such leakage 
is one of the most difficult and costly types of errors to detect and address, making it unfavorable to encode qubits in large multi-level systems such as neutral atoms. 
Our method to
address these leakage errors makes use of techniques based on optical pumping, such that the multi-level structure of each atom can be utilized as part of the redundancy required for QEC. While we focus on neutral atom-based quantum information processors, these techniques are adaptable to many other hardware platforms---for example, they could also greatly facilitate the correction of leakage-type errors in superconducting qubits or trapped ions. For the Rydberg-atom systems we study, we design a method that even converts all leading-order errors to Pauli-$Z$ type errors (Figure \ref{fig:overview-figure}c), which then allows us to develop particularly efficient FTQC protocols.

The  manuscript is organized as follows: 
we begin in Section \ref{sec:summary} by outlining the  key insights and main results of this work. A detailed analysis of the error channels in the Rydberg system is presented in Section \ref{sec:error-models}.  Under this realistic  error model, we design FTQC schemes based on the seven-qubit Steane code in Section \ref{sec:ftqc-7}. Furthermore, by utilizing atomic species with high nuclear spin, we develop an alternative, leading-order fault-tolerant protocol in Section \ref{sec:ftqc-repetition} based on a simple repetition code. We then show in Section \ref{sec:experiments} how the key ingredients of our proposals can be implemented in near-term experiments. Finally, we present conclusions and outlook in Section \ref{sec:conclusions}. 

\section{Overview of Main Results}
\label{sec:summary}

\begin{figure}[t]
\includegraphics[width=\textwidth]{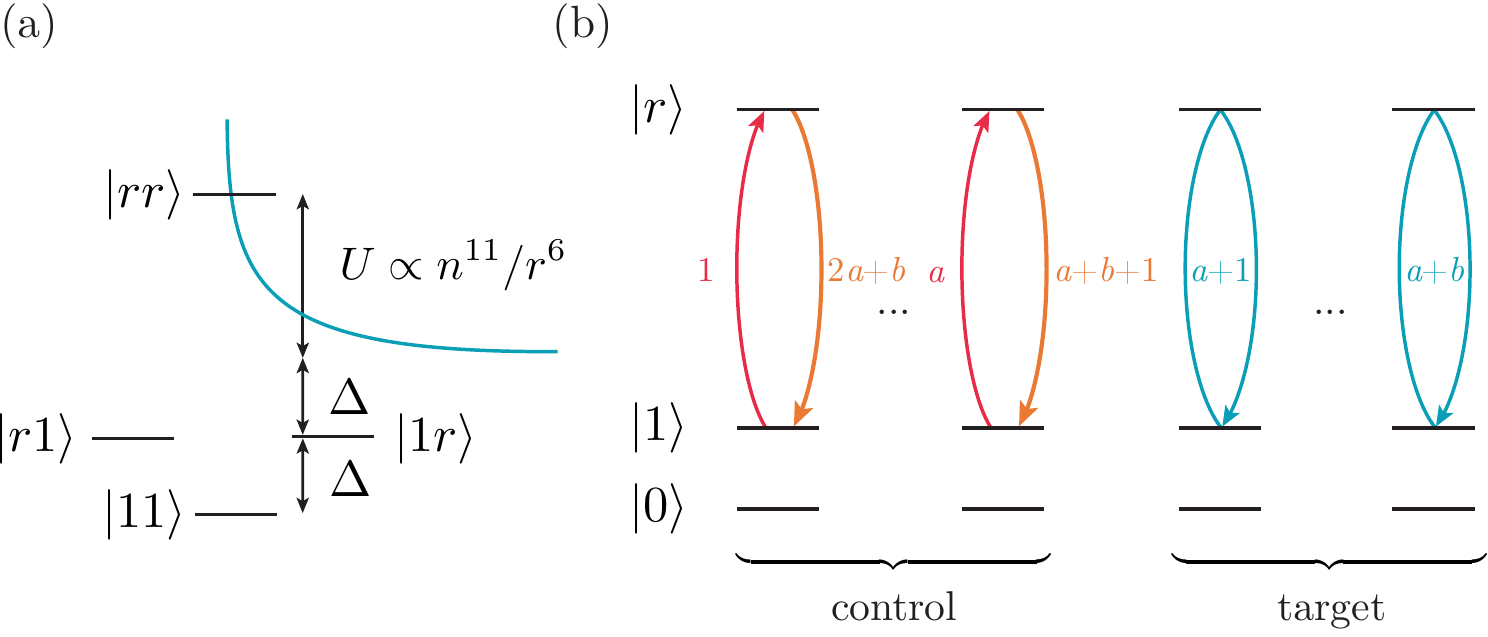}
\caption{(a) Rydberg blockade mechanism. $\Delta$ is the Rydberg laser detuning, and the Rydberg interaction strength $U \propto n^{11} /r^6$, where $n$ is the principal quantum number and $r$ is the atom separation. (b) Protocol for performing a multi-qubit entangling Rydberg gate $R(C_1, C_2, ..., C_a; T_1, T_2, ... T_b)$ on a set of atoms which are all within one given blockade volume. Resonant $\pi$ pulses $|1\rangle \leftrightarrow |r\rangle$ are first applied to each control qubit (red arrows), followed by $2\pi$ pulses on each target qubit (blue arrows). The control qubits are then returned to the ground state manifold via the $\pi$ pulses shown in orange. 
Labels on the arrows indicate the ordering of pulses. This Rydberg gate is related to the more conventional controlled-phase gate $\textrm{C}^a \textrm{Z}^b$ by conjugating all control qubits and one target qubit by Pauli-$X$ operations, or by applying Pauli-$Z$ gates on both control and target qubits in the special case of $a=b=1$ ($\textrm{CZ} = R(C_1;T_1) Z_{C_1}Z_{T_1}$). It can also be used to implement $\textrm{C}^a\textrm{NOT}^{b}$ from $\textrm{C}^a \textrm{Z}^b$ by conjugating the target qubits by Hadamard gates. The Rydberg gate $R(C_1, C_2, ..., C_a; T_1, T_2, ... T_b)$ is sometimes referred to as a ``collective gate.''}
\label{fig:collective-gates}
\end{figure}

We consider neutral atoms in a static magnetic field $\mathbf{B} = B_z \hat{\mathbf{z}}$. Due to the nonzero nuclear spin $I$, the electronic ground state manifold consists of many sub-levels split by hyperfine coupling and a finite $\mathbf{B}$ field. These levels  exhibit remarkably long lifetimes, making them particularly good candidates for encoding qubits (or more generally, qudits) for  quantum information processing. Furthermore, although neutral atoms in ground electronic states are effectively non-interacting, entangling gates between nearby atoms can be performed by coupling one of the qubit states (e.g.\ $|1\rangle$) to a  Rydberg $nS$ state $|r\rangle$ with large $n$, which exhibits strong van der Waals interactions (Figure \ref{fig:collective-gates}a). Under certain conditions, these interactions can be interpreted effectively as a blockade constraint prohibiting simultaneous Rydberg population within a blockade radius $R_B$. These can be leveraged to perform, for example, fast multi-control, multi-target phase gates 
$R(C_1, C_2, ..., C_a; T_1, T_2, ... T_b)$ (sometimes also referred to as ``collective gates''), which are related to the standard $\textrm{C}^a\textrm{Z}^{b}$ gates upon conjugating all control qubits $C_j$ and the first target qubit $T_1$ by Pauli-$X$ gates~\cite{Jaksch00, Isenhower11,Levine19};
this is achieved by applying individually addressed, resonant $\pi$ and $2 \pi$ pulses between the qubit $|1 \rangle$ state and the Rydberg state (Figure \ref{fig:collective-gates}b). Such an operation is also related to the gate $\textrm{C}^a\textrm{NOT}^{b}$ by single-qubit unitaries and has been demonstrated in recent experiments for small $a,b$ \cite{Levine19}.

While this procedure provides an efficient scheme to entangle  two or several atoms, for large-scale quantum computations, the finite lifetime of Rydberg states presents an important  source of error even if the rest of the experimental setup is perfect. This lifetime is determined by several contributions. First, interactions with blackbody photons can induce transitions from the $nS$ state to nearby Rydberg $n'P$ states of higher or lower energy; such errors are subsequently referred to as blackbody radiation-induced (BBR) errors. Second, spontaneous emission of an optical frequency photon can result in radiative decay (RD) to a low-lying $P$ state, which will quickly relax into the ground state manifold. 
In addition, if a multi-photon Rydberg excitation scheme is used for the Rydberg pulses, another intrinsic source of error during Rydberg gates is photon scattering from an intermediate state. These error channels are illustrated in Figure \ref{fig:overview-figure}c.

For the purposes of QEC, these errors can be formally described as follows 
(see Section
\ref{sec:error-models}):  BBR errors give rise to quantum jumps from the qubit $\ket{1}$ state to Rydberg $P$ states (corresponding to a leakage error), as well as Pauli-$Z$ errors within the qubit manifold, while RD and intermediate state scattering may also result in  quantum jumps from $|1 \rangle$ to the Rydberg $nS$ state or other hyperfine ground states. The relative error probabilities are determined by selection rules and branching ratios. In addition to these intrinsic errors, we also study the errors in the experimental setup such as Rydberg pulse imperfections or finite atomic temperature. We find that these experimental errors fall within a subset of the RD error model and can therefore also be addressed using our techniques. We note that, throughout this work, we assume the rotations within the hyperfine manifold have much higher fidelity than the Rydberg pulses, as is typically the case. Such errors can also be suppressed to high orders by using existing experimental methods such as composite pulse sequences or by incorporating traditional QEC techniques such as concatenation.

\subsection{Reduction to Pauli-$Z$ errors}

To protect against the errors mentioned above, 
three critical observations are used (see Figure \ref{fig:overview-figure}c).  First, we note that quantum jumps from $|1 \rangle$ to Rydberg states associated with BBR can be detected via the Rydberg blockade effect by using a nearby ancilla qubit, and subsequently converted to a Pauli-$Z$ type error by ejecting the Rydberg atom and replacing it with a fresh atom prepared in the $|1\rangle$ state~\cite{Bluvstein21b}. Second, quantum jumps from $|1 \rangle$ to ground state sublevels outside the qubit subspace can be corrected via optical pumping techniques. This is particularly efficient as it does not require any qubit measurement for feed-forward corrections, unlike traditional proposals for correcting leakage errors~\cite{Auger17}. 
Third, for atomic species with large enough nuclear spin, dipole selection rules prevent a stretched Rydberg state from decaying to certain ground state sublevels. By making use of this multi-level structure of neutral atoms along with the high-fidelity manipulations of hyperfine states, we can ensure that RD and intermediate state scattering errors do not result in $|1 \rangle \rightarrow |0 \rangle$ transitions, thereby eliminating $X$ and $Y$ type errors from the error model. This reduction of error types can significantly alleviate the resource requirement for FTQC. 

\begin{table*}[t]\caption{Comparison of resource costs for fault-tolerant measurement of all stabilizers to correct Pauli errors to first order. Numbers in parentheses indicate the maximum number of operations required in the unlikely scenario where an error is detected. Details on how to obtain the gate counts for the Ryd-7 and Ryd-3 protocols can be found in Appendix~\ref{sec:resource-cost-app}.}\label{tab:stab-comparison}
\begin{tabular}{|c|c|c|c|c|}
\hline & {\sc 2-qubit gates} & {\sc 3-qubit gates} & {\sc Ancillas} \\ \hline
{\sc 7-qubit flagged \cite{Chao18}}  & 36 (48) & 0 & 2 \\ \hline
{\sc 15-qubit flagged \cite{Chao18}} & 80 (112) & 0 & 2 \\ \hline
\bfseries{\scshape{Ryd-7}} & {\bf 24 (36)} & {\bf 0} & {\bf 2}  \\ \hline
\bfseries{\scshape{Ryd-3}} & {\bf 8 (16)} & {\bf 4 (8)} & {\bf 4}  \\ \hline
\end{tabular}
\end{table*}

\begin{table*}[t]\caption{Comparison of resource costs for the highest-cost fault-tolerant logical operation. $\CCZ$ denotes the three-qubit controlled-controlled-phase gate, while $H$ denotes the single-qubit Hadamard gate. Numbers in parentheses indicate the maximum number of operations required in the unlikely scenario where an error is detected. For the Rydberg protocols, the gate counts presented assume a blockade radius of $3d$, where $d$ is the nearest-neighbor lattice spacing. Derivations of the gate counts for the Ryd-7 and Ryd-3 protocols can be found in Appendix~\ref{sec:resource-cost-app}, while details on how to obtain the blockade radius requirement can be found in Appendix~\ref{sec:blockade-app}.}\label{tab:gate-comparison}
\begin{tabular}{|c|c|c|c|}
\hline  & {\sc 2-qubit gates} & {\sc 3-qubit gates} & {\sc Ancillas} \\  \hline
{\sc Yoder, Takagi, and Chuang \cite{Yoder16} (CCZ)} & 162 & 21 & 72   \\ \hline
{\sc Chao and Reichardt \cite{Chao18b} (CCZ)} & 1352 (1416) & 84 & 4  \\ \hline
\bfseries{\scshape{Ryd-7 (CCZ)}} & {\bf 0 (78)} & {\bf 27 (29)} & {\bf 2} \\ \hline
\bfseries{\scshape{Ryd-3 (CCZ)}} & {\bf 0 (18)} & {\bf 27 (27)} & {\bf 4} \\ \hline
\bfseries{\scshape{Ryd-3 (H)}} & {\bf 20 (28)} & {\bf 53 (57)} & {\bf 10} \\ \hline
\end{tabular}
\end{table*}

\subsection{Fault-tolerant protocols}

We now describe two FTQC protocols to address these intrinsic errors in neutral Rydberg atom platforms. The first is based on the seven-qubit Steane code~\cite{Steane96}, while the second uses the three-qubit repetition code; the latter is more compact and efficient, but has additional experimental requirements such as control over multiple Rydberg states and more complex encoding of logical operations. 
To realize the seven-qubit code (Ryd-7), we notice that logical state preparation, stabilizer measurements, and a universal set of logical gates (Hadamard and Toffoli \cite{Shi03}) can be implemented using only controlled-phase ($\CZ$) or controlled-controlled-phase ($\CCZ$) gates, up to single-qubit unitaries at the beginning and end of the operation. For example, while the stabilizer measurements are typically presented as a sequence of CNOT gates between the data atoms and an ancilla atom, these CNOT gates can be constructed by conjugating a $\CZ$ gate with Hadamard gates on the target qubit.  By mapping each Rydberg gate error to a Pauli-$Z$ error, we therefore ensure that it will commute with all subsequent entangling gates in the logical operation or stabilizer measurement, so it does not spread to other qubits (Figure \ref{fig:overview-figure}b). The resulting single-qubit $X$ or $Z$ error can be corrected by the seven-qubit code in a subsequent round of QEC. This eliminates the need for ``flag qubits,'' which are otherwise necessary to prevent spreading of errors as discussed in Ref.~\cite{Chao18,Chao18b}. To further reduce resource costs for experimental implementation, we make additional use of the structure of the Rydberg error model, stabilizer measurement circuits, and logical operations of the seven-qubit code. For instance, one of our key findings is that leakage errors into other Rydberg states do not need to be corrected after every Rydberg gate, but can be postponed to the end of a stabilizer measurement (e.g.\ Figure~\ref{fig:overview-figure}b). This allows us to minimize the number of intermediate measurements necessary for each FTQC component, which is typically a limiting factor in state-of-the-art neutral atom experiments.

The simplified error model introduced by conversion of all Rydberg gate errors to Pauli-$Z$ errors  motivates us to use the three-qubit repetition code instead of the seven-qubit code to design a leading-order fault-tolerant protocol (Ryd-3). In this case, 
the stabilizer measurement circuits are also comprised of CNOT gates on data atoms controlled by the ancilla. However, the implementation of each CNOT must be modified: when a $\CZ$ gate is conjugated by Hadamard gates as in Figure \ref{fig:overview-figure}b, 
a Pauli-$Z$ type error that occurs during the CZ gate will be converted to a Pauli-$X$ error after the Hadamard. Such an error can no longer be corrected by the repetition code. 
Additional errors, such as radiative decay of a control qubit prior to manipulation of the target qubit, can lead to error spreading and correlated errors.

These errors can be addressed via 
a protocol to directly implement CNOT gates in a {\it bias-preserving} way, such that these implementations will not generate any Pauli-$X$ and $Y$ errors to leading order (Figures \ref{fig:bias-preserving-cnot-target} and \ref{fig:bias-preserving-cnot-control}). Our protocol makes use of the rich multilevel structure of atoms with large nuclear spin ($I \ge 5/2$, e.g.~$^{85}$Rb, $^{133}$Cs, $^{87}$Sr, \ldots), as well as additional Rydberg states for shelving. Furthermore, we leverage the fact that pulses between certain (i.e.\ hyperfine) levels can be performed with very high fidelity, so that leading-order errors involve only Rydberg state decay or Rydberg pulse imperfections. This assumption is particularly important, as Ref.~\cite{Guillaud19} shows a no-go theorem stating that a bias-preserving CNOT gate cannot be implemented in any qudit system with a finite number of levels without such structure in the error model. To circumvent this, our pulse sequence directly implements a hyperfine Pauli-$X$ gate on the target qubit only if a nearby Rydberg atom is present (without the need for subsequent Hadamard gates), and we show that errors during this sequence can all be mapped to Pauli-$Z$ errors. Additionally, correlated errors due to control-atom decay can be prevented by using multiple control atoms, such that if one atom decays, the remaining atom(s) still ensure proper gate operation on the target atom.  This bias-preserving CNOT protocol can be directly generalized to implement a bias-preserving Toffoli operation, enabling a leading-order fault-tolerant implementation of each operation of the three-atom repetition code.  
Throughout the manuscript, we 
use the term ``leading-order fault-tolerance'' in referring to the Ryd-3 protocol as our framework does not inherently address all single-qubit errors, but existing experimental techniques such as composite pulse sequences can be used in conjunction with our protocol to suppress such errors to higher orders (see Section~\ref{sec:other-errors}).

Upon comparing our protocols with existing, general-purpose FTQC proposals, we find that
the number of required physical qubits and gates for both of our approaches is dramatically reduced (Tables \ref{tab:stab-comparison}, \ref{tab:gate-comparison}). For example, as seen in Table~\ref{tab:gate-comparison}, performing the highest-cost operation from our logical gate set, our Ryd-7 protocol requires only 2 ancilla qubits compared with 72 ancillas  in Yoder, Takagi, and Chuang~\cite{Yoder16}. Likewise, Ryd-7 uses at most 60 2-qubit gates (when errors are detected) to perform this logical operation, instead of 1416 gates as in Chao and Reichardt~\cite{Chao18b}. Such a significant reduction is possible for our protocols because we leverage both the special structure of the error model and the unique capabilities of Rydberg setups.

Several aspects of  
the comparison above should be considered. Specifically, we note
that certain single-qubit errors addressed in Refs.~\cite{Yoder16, Chao18b} cannot be corrected in our protocols (e.g. Pauli-$X$ errors arising from rotations in the hyperfine manifold). However,  we emphasize that Refs.~\cite{Yoder16, Chao18b} also did not consider additional types of errors such as leakage errors which are corrected by our protocol. Indeed, incorporating leakage correction would further increase the resource cost for the earlier  proposals considerably. As such, Tables~\ref{tab:stab-comparison}, \ref{tab:gate-comparison} must be interpreted as a comparison of the cost ensuring fault-tolerance against the leading-order sources of error in a given setup. In the case of Refs.~\cite{Yoder16, Chao18b}, these errors include all single-qubit Pauli errors but not leakage errors, while in Rydberg systems, one must address leakage errors at leading order but can neglect certain single-qubit errors.

\subsection{Towards experimental implementation}

For scalable implementation of our FTQC protocols, it is important to consider the geometrical placement of atoms. In addition, because Rydberg entangling gates can only be implemented between atoms within the blockade radius $R_B$, each protocol defines a minimum value of $R_B$ (in units of $d$, which is the smallest atom-atom separation). We find that both the Ryd-7 and Ryd-3 protocols can be implemented naturally when the atoms are placed on the vertices of a triangular lattice as shown in Figure~\ref{fig:overview-figure}a,d. For both protocols, the required Rydberg gates can be implemented when the blockade radius ($R_B$ for Ryd-7, or the larger radius $R_{B,1}$ for Ryd-3) is greater than $3d$, an interaction range which has already been demonstrated in recent experiments~\cite{Bernien17}. This requirement can be further reduced in both cases if it is possible to move atoms in between certain operations while preserving the coherence of hyperfine ground states, a capability which has been recently realized~\cite{Bluvstein21b}.

Each component of our FTQC schemes can be implemented in near-term experiments. For neutral alkali-atom systems, recent experiments have already achieved high-fidelity control and entanglement leading to remarkable demonstrations of quantum simulations and computations~\cite{Levine18,Levine19,Saffman10}. The near-deterministic loading of atoms into lattice structures as shown in Figure \ref{fig:overview-figure} has already been realized in two and three dimensions~\cite{Barredo16,Ebadi20,Barredo18,Barredo20}.

To perform QEC in our protocol, an important requirement is the ability to measure individual qubits and/or detect Rydberg population while preserving coherence in nearby atoms, such that  feed-forward correction can be performed. 
Several approaches for performing fast measurements of individual qubit states in selected 
atoms can be realized. For example, these selected atoms can be moved into a ``readout zone'' where their qubit state can be rapidly detected via fast, resonant photon scattering on a cycling transition. Alternatively, one could use arrays with two species (such as two isotopes of the same atom or two different atomic species), where the data atoms are encoded in one atomic species and ancilla atoms are encoded in another species that can be easily measured~\cite{Zeng17,Singh21}.  Finally, the fast detection of Rydberg states has been recently demonstrated in small atomic ensembles using Rydberg electromagnetically induced transparency (EIT)~\cite{Xu21}. These  could be integrated with the tweezer array platforms currently used for quantum information processing. In these EIT-based procedures, the Rydberg blockade effect translates to clean signatures in the absorption spectrum, and the collectively enhanced Rabi frequency allows for ultrafast detection on a microsecond time scale~\cite{Xu21}.

While we focus primarily on neutral alkali atoms in this work, significant developments have also been made using alkaline-earth atoms for Rydberg-based quantum computations \cite{Madjarov20,Wilson19}. The clock transition in these atoms allows for high-fidelity qubit encodings, and the large nuclear spin in fermionic species is particularly advantageous for our protocols, so we conclude by discussing how our FTQC schemes can be generalized and applied to these experiments.  More detailed experimental considerations are discussed in Section \ref{sec:experiments}.

\section{Error Channels in Rydberg Atoms}
\label{sec:error-models}

In this section, we analyze dominant error mechanisms for quantum operations involving Rydberg atoms (Figure~\ref{fig:overview-figure}c). Because the predominant errors in single-qubit operations can be suppressed to high orders via composite pulse sequences~\cite{Vandersypen05,Bluvstein21b}, we may primarily focus on errors occurring during Rydberg-mediated entangling operations.
The decay channels of the Rydberg states include blackbody radiation-induced (BBR) transitions and spontaneous radiative decay (RD) transitions to lower-lying states~\cite{Beterov09}. 
Depending on the specific choice of atomic species, another  source of error for Rydberg gates can be the scattering from an intermediate state if a two- or multi-photon excitation scheme is used; this is the case for excitation of $^{87}$Rb or $^{85}$Rb to Rydberg $nS$ states \cite{Bernien17}. 
We will assume these effects are the predominant source of errors that occur during the entangling operations, and we consider contributions to the error model to leading order in the total error probability.

\subsection{Error modeling for BBR transitions}
\label{sec:bbr}

When a BBR transition occurs on one of the atoms during an entangling gate, it signals that this  atom has started 
in the $|1 \rangle$ state, since $|0 \rangle$ is not coupled to $|r \rangle$. Such a procedure corresponds to a `quantum jump' as discussed in, for example, Ref.~\cite{Carmichael93}. The resulting state will predominantly be a nearby Rydberg state $|r' \rangle$ compatible with dipole selection rules. Due to the relatively long lifetimes of Rydberg states, we may assume that the atom will not decay again within the timescale of several Rydberg gate operations, as these would be higher-order processes. In this case, because the states $|r' \rangle$ are not de-excited in the ensuing operations, one serious consequence of BBR quantum jumps is that the remaining Rydberg operations on atoms within the interaction range will be affected by blockade, potentially resulting in multiple, correlated Pauli-$Z$ type errors. Less intuitively, even if a quantum jump does not occur during the gate operation, the atom's state is still modified due to evolution under a non-Hermitian Hamiltonian: it will be more likely that the atom started out in the $|0 \rangle$ state. More details on the theory of quantum jumps can be found in Ref.~\cite{Carmichael93}.

For the purposes of QEC, it is useful to express the decay channels in the Kraus operator form, where time evolution of a density operator is given by $\rho \mapsto \sum_\alpha M_\alpha \rho M_\alpha^\dagger$ 
and the {\it Kraus operators} $M_\alpha$ satisfy the completeness relation $\sum_\alpha M_\alpha^\dagger M_\alpha = \mathbf{1}$ \cite{Preskill98}. For the BBR error model, there will be one Kraus operator
\begin{equation}
    M_{{r'}} \propto |r' \rangle \langle 1|
\end{equation}
for each possible final Rydberg state $\ket{r'}$, where the proportionality constant is determined by the BBR transition rate from $\ket{r}$ to $\ket{r'}$ (see Appendix \ref{sec:transition-rates-app}). In the absence of quantum jumps, the evolution is given by the Kraus map
\begin{equation}
M_0 = \sqrt{1-P} |1 \rangle \langle 1| + \sum_{\ket{n} \neq \ket{1}} |n \rangle \langle n|,
\end{equation}
where $P$ is the probability for a BBR transition to occur. 

During entangling operations, these BBR errors can give rise to correlated errors. For example, in the Rydberg gates shown in Figure~\ref{fig:collective-gates}, a target qubit can only incur a BBR error if the control qubits were all in the $|0 \rangle$ state. Thus, for the $\textrm{C}^a \textrm{Z}^b$ gates shown in Figure~\ref{fig:collective-gates}, the possible correlated errors may involve one of the Kraus maps $M_{r'}$ or $M_0$ occurring on one of the qubits, together with $Z$-type errors on some or all of the remaining qubits involved in that gate. 

The rate of BBR transitions from a given Rydberg state $nL$ to another specific state $n'L'$ can be calculated from the Planck distribution of photons at the given temperature $T$ and the Einstein coefficient for the corresponding transition (see Ref.~\cite{Beterov09}). For $^{87}$Rb atoms excited to the $70S$ Rydberg state, there are four dominant final states associated to these BBR errors (see Appendix \ref{sec:transition-rates-app}); these are illustrated in Figure~\ref{fig:overview-figure}c as red arrows. 
The total rate of BBR transitions summed over all possible final states is \cite{Cooke80}
\begin{equation}
\label{eq:bbr-scaling}
\Gamma_{\textrm{BBR}} = \frac{4k_B T}{3c^3 n_{\textrm{eff}}^2},
\end{equation}
where $k_B$ is Boltzmann's constant, $c$ is the speed of light, and $n_{\textrm{eff}}$ is the effective principal quantum number of the Rydberg state which determines its energy \cite{Saffman10}: $E_{nL} \propto -1/(2n_{\textrm{eff}}^2)$. We note that the overall rate of BBR transitions can be suppressed by operating at higher $n_\textrm{eff}$ or operating at cryogenic temperatures. 

\subsection{Error modeling for RD transitions}
\label{sec:radiative-decay}

The spontaneous emission events corresponding to RD transitions can be modeled as quantum jumps involving the emission of an optical-wavelength photon. Unlike BBR, however, the resulting state will be a low-lying $P$ state, which will quickly decay back into the ground state manifold. For the stretched Rydberg state of $^{87}$Rb, the RD transitions are almost entirely two- or four-photon decay processes to one of the five states in the ground state manifold indicated by light blue arrows in Figure~\ref{fig:overview-figure}c (see Appendix \ref{sec:transition-rates-app} for the precise branching ratios). For the purpose of QEC, we will separately consider the cases of decay into the qubit $|1 \rangle$ state and decay into one of the other ground state sub-levels. Because the spontaneous emission event can occur anytime during the Rydberg laser pulse, the first type of decay can result in a final state which is a superposition of $\ket{1}$ and $\ket{r}$. Upon averaging over all possible decay times during the entire pulse (see Appendix \ref{sec:master-eq-app}), one finds that these errors can be modeled using a combination of $Z$-type errors and leakage into the $\ket{r}$ state, with the Kraus operators 
\begin{equation}
\begin{gathered}
M_0 = |r \rangle \langle r| + \alpha |1 \rangle \langle 1| + \beta |0 \rangle \langle 0|, \\
M_{r} \propto |r \rangle \langle 1|, \quad
M_{1} \propto |1 \rangle \langle 1|, \quad 
M_{2} \propto  |0 \rangle \langle 0|,
\end{gathered}
\end{equation}
where $\alpha$, $\beta$, and the proportionality constants depend on the probability for the atom to incur an RD transition to the $\ket{1}$ state and the specific Rydberg pulse being performed.

At the same time,  decay to one of the other ground state sublevels shown in Figure~\ref{fig:overview-figure}c leads to leakage out of the computational subspace as in the traditional QEC setting (without influencing Rydberg operations on neighboring atoms). That is, for each hyperfine state $\ket{f} \neq \ket{1}$, we have a Kraus operator
\begin{equation}
M_{f} \propto |f \rangle \langle 1|,
\end{equation}
where the proportionality constant depends on the probability for an RD transition and the branching ratio from $|r \rangle$ to the specific state $|f \rangle$ (see Appendix~\ref{sec:transition-rates-app}). Note that due to dipole selection rules, the number of RD channels with non-negligible final state probability is minimized by choosing to couple the $|1 \rangle$ state to a so-called ``stretched Rydberg state'' for entangling gates
\footnote{
If, for example, we had instead chosen a Rydberg state with $m_J+m_I=0$, there would be several additional 
final states in each case.}.
In particular, in this analysis, the decay into the qubit $\ket{0}$ state is negligible to leading order. Such an event, corresponding to the Kraus operator $M \propto |0 \rangle \langle 1|$ (or equivalently, Pauli-$X$ and $Y$ errors), is considered when we discuss methods to suppress residual errors in our protocols. 

As in the BBR case, the absence of quantum jumps results in the atom's population being shifted toward the $\ket{0}$ state, which can be modeled using Pauli-$Z$ errors. RD errors can also give rise to correlated errors when they occur during the primitive entangling gates illustrated in Figure~\ref{fig:collective-gates}. In this case, possible correlated errors may involve one of the aforementioned Kraus maps occurring on one of the qubits, together with Pauli-$Z$ and/or $|r \rangle \langle 1|$ errors on some or all of the remaining qubits involved in that gate.

While as noted above, the rate of BBR transitions depends upon the temperature $T$ and $n_\textrm{eff}$, the total RD rate is temperature-independent. 
Due to reduced overlap between the atomic orbitals, it scales as $\Gamma_0 \sim 1/n_\textrm{eff}^3$ \cite{Low12}. Comparing this with the scaling for the BBR decay rate, we see that while both error rates decrease for larger $n$, BBR processes dominate for large $n$, and RD processes dominate for smaller $n$ or very low $T$. 

\subsection{Errors from intermediate state scattering}
\label{sec:is-scattering}

When multi-photon excitation is used to couple the $\ket{1}$ state to the Rydberg state, scattering from an intermediate state can give rise to another important intrinsic source of error. By using $\sigma^+$-polarized light in the first step of the excitation and choosing the intermediate state to be a $P_{3/2}$ state with the lowest possible $n$, the intermediate state scattering channels form a subset of the RD channels---they can only result in decay into the qubit $\ket{1}$ state or two other hyperfine ground states, as shown in grey in Figure~\ref{fig:overview-figure}c~\footnote{If an intermediate state with higher $n$ is used, such as $6P_{3/2}$, this is still true to leading order; however, a (highly improbable) four-photon process could potentially lead to mixing between the qubit states $|1\rangle$ and $|0\rangle$.}. Thus, whenever intermediate state scattering is not explicitly mentioned in the following sections, we will assume it has been incorporated with RD errors. We also note that this error rate can be suppressed by increasing intermediate laser detuning in the multi-photon transition, while also increasing laser power. 

\subsection{Experimental imperfections}
\label{sec:other-errors}

While BBR, RD, and intermediate state scattering processes constitute the dominant errors for Rydberg-mediated collective gates, it is also important to consider other forms of error, such as technical imperfections in the experimental setup.
As discussed in Refs.~\cite{Bernien17,Levine18,Bluvstein21b}, the most significant errors of this kind are atom loss and fluctuations in laser phase, intensity, and frequency. The 
Rydberg laser fluctuations can all be modeled using Pauli-$Z$ errors and leakage into the $\ket{r}$ state, so these errors can be addressed together with the other errors discussed above. Finite atomic temperature, resulting in velocity spread and Doppler broadening on the Rydberg transition~\cite{Levine18}, likewise leads to Pauli-$Z$ errors and leakage into the $\ket{r}$ state. Temperature-induced positional spread causes similar errors, and due to the robustness of the blockade-based gate, these errors can even be rendered negligible with sufficiently large interaction strengths~\cite{Levine19}. On the other hand, atom loss forms a more complicated version of a leakage error (called {\it erasure} in the quantum information literature~\cite{Gottesman10}). However, as discussed in Appendix~\ref{sec:atom-loss-app}, we find that such errors can also be addressed efficiently in the present framework. In certain cases, the special properties of these errors can be further leveraged to improve QEC efficiency, as done in the recent proposal of Ref.~\cite{Wu22}.

Experimental imperfections can also affect the hyperfine qubits used for storing quantum information and performing single-qubit gates; however, these primarily result in Pauli-$Z$ errors and leakage to other hyperfine states, which group together with the error types described above. Moreover, these tend to be significantly smaller sources of error than the two-qubit gates~\cite{Bluvstein21b}. By choosing a magnetically insensitive transition for our qubit states, we eliminate the leading order errors arising from magnetic field fluctuations. However, 
$Z$-type dephasing errors can still arise from the differential light shift from the optical trap. Finite atomic temperature, fluctuating tweezer power, and atom heating can thus cause dephasing, although these can be alleviated to achieve qubit coherence times $T_2 \sim 1$~s by applying standard dynamical decoupling sequences~\cite{Bluvstein21b}. Leakage to other hyperfine $m_F$ states can also occur due to so-called Raman scattering from the tweezer light, but these effects can be greatly suppressed to timescales $> 10$~s by sufficiently detuning the tweezer light~\cite{Bluvstein21b}. Since our qubit states are separated by $\Delta m_F = 2$ (a nuclear-spin-flip transition), bit-flip $X$ and $Y$ error rates from tweezer-induced scattering are even smaller. Finally, temperature-induced Doppler effects, which could in principle result in $Z$-type errors, are negligible since the qubit transition is of microwave-frequency, and microwave phase stability can be exceptional on the Raman laser used for single-qubit manipulations.

At the same time, as noted earlier, certain experimental imperfections associated with the hyperfine rotations are not directly corrected with our protocol, but can be minimized or suppressed via other mechanisms such as composite pulse sequences. For example, the primary source of single-qubit gate errors in recent experiments involves laser amplitude drifts or pulse miscalibrations, which can result in $X,Y$, and $Z$-type errors~\cite{Bluvstein21b}. 
However, these coherent errors can be significantly suppressed by using composite pulse sequences, as done in Ref.~\cite{Bluvstein21b}: in particular, the BB1 pulse sequence suppresses pulse amplitude errors to sixth order~\cite{Vandersypen05}. 
On the other hand, the error rates associated with phase noise in single-qubit gates are typically much smaller: for example, the phase noise in $^{171}$Yb$^+$ hyperfine qubits has been shown to limit coherence to order 5000 seconds~\cite{Wang21}. Although other sources of frequency flucutations result in a $T_2^*$ of approximately 4~ms for the Rb qubit of Ref.~\cite{Bluvstein21b}, thereby inducing pulse frequency errors,  these errors are strongly suppressed to second-order due to the MHz-scale Raman Rabi frequencies, and they can be further suppressed with improved cooling and microwave source stability. Furthermore, they can be made completely negligible by using appropriate composite pulse sequences~\cite{Vandersypen05}.
Finally, incoherent scattering from the Raman beams used for single-qubit rotations can also cause leakage and $X,Y$-type errors, which can be on the $10^{-5}$ level~\cite{Bluvstein21b} for far-detuned Raman beams used for electron-spin-flip transitions but may be higher for nuclear-spin-flip transitions as used for the qubit states here. These remaining hyperfine qubit error rates are significantly smaller than the primary sources of error considered, and they can be further corrected via concatenation of additional error correction codes. 

\subsection{Summary of error channels}

We have shown that the 
multi-level nature of neutral atoms gives rise to various complexities in the error model, including a large number of decay channels and the possibility for Rydberg leakage errors to influence many future operations, resulting in high-weight correlated errors. Despite these complications, one important feature of our error model makes it substantially simpler than the set of all Pauli errors studied in more generic setups---no Pauli-$X$ or $Y$-type errors are introduced during our Rydberg gates. Indeed, in the following sections, we will show how all the additional leakage errors and correlated errors in our error model can be converted into $Z$-type errors, and we use this to design FTQC protocols with substantially reduced resource costs. This reduction to Pauli-$Z$ errors can be found in Sections \ref{sec:ryd-bbr}, \ref{sec:ftqc-lf} for the seven-qubit code and Section~\ref{sec:rydberg-bias-preserving} for the repetition code.

\section{FTQC with the Seven-Qubit Steane Code}
\label{sec:ftqc-7}

Having established the error model for the Rydberg operations, we now proceed to develop fault-tolerant schemes to detect and correct these errors and perform a universal set of logical operations. The key concept for this construction is the ability to convert all errors described in the previous section into Pauli-$Z$ type errors by introducing ancilla qubits and using the blockade effect, dipole selection rules, and optical pumping (see Figure~\ref{fig:overview-figure}c). We begin by demonstrating the protocol when only BBR errors are significant (i.e.\ in the limit of higher Rydberg principal quantum number $n$), as the error model and QEC mechanisms are simpler to understand in this case. The universal gate set we develop comprises a logical Hadamard gate and a logical controlled-controlled-phase ($\CCZ$) or Toffoli gate \cite{Shi03}. We then describe the more general case involving both BBR and RD errors. Subsequently, we compare the resource cost of our protocol against other fault-tolerant computation schemes and discuss considerations for scalable computation. The final scheme we present in this section is referred to as Ryd-7. Throughout this section, we will use qubits encoded in $^{87}$Rb as a concrete example to illustrate our protocols.

While various equivalent definitions of FTQC have been given in the literature for traditional error models, to accommodate the possibility of Rydberg leakage errors---that is, any Rydberg population remaining after the gate operation---we must use the following, stricter one:
\begin{definition*}
A distance-$d$ QEC code is {\it fault-tolerant} if after any round of error detection and correction, to order $\left(p_{\textrm{tot}}\right)^t$, at most $t$ single-qubit Pauli errors are present, where $t = \left\lfloor\frac{d-1}{2}\right\rfloor$ and $p_{\textrm{tot}}$ is the sum of all error probabilities. In addition, no Rydberg population can be present after any round of error detection and correction.
\end{definition*} 
The final requirement is important because any remnant Rydberg population could blockade future Rydberg gates.

In the following, we will examine the case of code distance $d = 3$ and $p_{\textrm{tot}} \sim O\left((\Gamma_{\textrm{BBR}}+\Gamma_0)/\Omega\right).$ Our QEC proposal has the following properties: to leading order in $p_{\textrm{tot}}$,
\begin{enumerate}
\item Code states can be prepared with at most a single physical qubit error, without leaving any final Rydberg state population.
\item After each round of error detection and correction, there is at most a single physical qubit error per logical qubit, and there is no  Rydberg state population.
\item Each logical gate introduces at most a single physical qubit error per involved logical qubit, without leaving any final Rydberg state population.
\end{enumerate}
It is straightforward to show that any distance-$3$ code satisfying the above properties is fault-tolerant.

Throughout the rest of the manuscript, we will use the term  {\it data qubit}  to refer to physical qubits used to encode a logical qubit, and {\it ancilla qubit} for physical qubits which are used to perform stabilizer measurements or detect errors.

\subsection{FTQC with BBR errors}
\label{sec:ryd-bbr}

\subsubsection{Qubit encoding}

Our quantum code is based on the seven-qubit Steane code, which uses a logical state encoding derived from classical binary Hamming codes \cite{Steane96}:
{\small
\begin{equation}
\begin{aligned}
\label{eq:steane-0}
\ket{0}_L = & \frac{1}{2\sqrt{2}}  (
 \ket{0000000} + \ket{1010101}+\ket{0110011}+\ket{1100110} \\
 & + \ket{0001111}+\ket{1011010}+\ket{0111100}+\ket{1101001})
\end{aligned}
\end{equation}
\begin{equation}
\begin{aligned}
\label{eq:steane-1}
\ket{1}_L = & \frac{1}{2\sqrt{2}}  ( \ket{1111111}+\ket{0101010}+\ket{1001100}+\ket{0011001} \\
& + \ket{1110000}+\ket{0100101}+\ket{1000011}+\ket{0010110}).
\end{aligned}
\end{equation}
}
The stabilizer operators for this code are
\begin{align}
\label{eq:steane-stabilizer}
&g_1 = IIIXXXX  & g_2& = IXXIIXX  & g_3 &= XIXIXIX \nonumber \\
&g_4 = IIIZZZZ & g_5& = IZZIIZZ  &g_6& = ZIZIZIZ.
\end{align}

In Eq.~(\ref{eq:steane-stabilizer}) and the rest of the manuscript when appropriate, we omit tensor product symbols and qubit indices and assume that the $j^{\textrm{th}}$ operator in each product acts on qubit $j$. Measurements of the stabilizers $g_1, ..., g_6$ allow for unique identification and correction of single-qubit $X$ and $Z$ errors. For instance, the absence of any error corresponds to all stabilizers $g_j = +1$, and a $Z$ error on the first qubit would be detected by $g_3 = -1$ and $g_j = +1$  for all $j \neq 3$. The error can then be corrected via an appropriate single-qubit gate. 

\subsubsection{Error detection and correction}
\label{sec:bbr-stabilizer-measurement}

To fault-tolerantly detect and correct for the errors associated with BBR events, we must be able to address both Rydberg leakage and Pauli-$Z$ errors. For the former case, even though leakage errors in traditional QEC settings can be particularly difficult to detect and correct, the particular form of leakage caused by BBR errors make them much easier to identify---we can use an ancilla and the blockade effect to detect the leaked Rydberg population. Specifically, we prepare a nearby ancilla qubit in the state  $\ket{+} = \frac{1}{\sqrt{2}}(\ket{0}+\ket{1})$ and apply a $2\pi$ Rydberg pulse to detect whether there is another Rydberg atom within the blockade radius. Due to the blockade effect, the ancilla will be in the $|+ \rangle$ (respectively, $|- \rangle$) state if nearby Rydberg population is (is not) present. 

Once detected, such errors can be easily converted to atom loss errors or $Z$-type errors. To convert the error to an atom loss error, we notice that the Rydberg atom naturally expels itself due to the anti-trapping potential of the tweezer~\cite{Bluvstein21b}, and can also be directly ejected in $\sim 100$~ns by pulsing a weak, ionizing electric field ($\sim$ 10~V/cm~\cite{Cohen03,Beguin13,Wu22}) which removes the ion and electron. The exact location of the ejected atom can be determined by following the atom loss protocol outlined in Appendix~\ref{sec:atom-loss-app} and Figure~\ref{fig:leakage-detection}; subsequently, the error can be corrected by replacing the ejected atom with a fresh atom prepared in the $|1\rangle$ state~\cite{Bluvstein21b} (thereby converting it to a $Z$-type error) and applying another round of QEC. To reduce the need for applying the atom loss correction protocol, one could add a preventative step after every entangling gate which incoherently re-pumps any remnant population in several most probable Rydberg states into the qubit $|1\rangle$ state. This procedure, along with more details on the conversion of Rydberg population errors, is further discussed in Appendix~\ref{sec:rydberg-leakage-app}.

For fault-tolerant error detection and correction, it is important to note that the ancilla used to probe for Rydberg population may also incur a BBR error.  This can be resolved by repeating the detection protocol upon finding a BBR error and also using a multi-step measurement procedure for the ancilla qubit; details are given in Appendix~\ref{sec:ancilla-multi-step-app}. Such a protocol will be assumed in all future sections when we use an ancilla to detect for Rydberg population.

To fault-tolerantly detect and correct for Pauli errors, we must measure the stabilizers (\ref{eq:steane-stabilizer}) in a manner robust against errors that may occur during the detection procedure. The stabilizers for this seven-qubit code are either products of Pauli-$X$ operators or products of Pauli-$Z$ operators, since the Steane code is a CSS code~\cite{Gottesman10}. The traditional (non-fault-tolerant) way to measure a product of four Pauli-$X$ operators (i.e.\ stabilizers $g_1$, $g_2$, or $g_3$) uses four controlled-phase gates conjugated by Hadamards (Figure~\ref{fig:overview-figure}b, black parts). 
Since Rydberg gate errors can occur during this protocol, we utilize a second ancilla qubit to detect for BBR errors after each entangling operation and convert them to $Z$-type errors when detected.

The $Z$ errors that occur during a Rydberg gate (or result from conversion of a BBR error) commute with the remaining $\CZ$ operations.  Thus, the only errors that can occur during a round of stabilizer measurements, to first order in $p_{\textrm{tot}}$, consist of a Pauli error acting on the ancilla and a Pauli error on one of the data qubits (Figure \ref{fig:overview-figure}b). By resetting the ancilla and repeating the measurement protocol when a $-1$ measurement outcome is obtained, we can eliminate the effect of the error on the ancilla qubit. An analogous method can be used for the $Z$ stabilizers. 
In this way, after each round of stabilizer measurements, the correct stabilizer eigenvalues can be obtained to leading order in $p_{\textrm{tot}}$, while introducing at most one physical qubit $X$ or $Z$ error.

While we have presented the fault-tolerant stabilizer measurement protocol in the simplest form where Rydberg state detection is performed after every physical gate, this is in fact not necessary. Indeed, if we postpone all such detection operations to the end of a circuit which measures the stabilizer $X_\alpha X_\beta X_\gamma X_\delta$ (where Rydberg gates are applied to data atoms in the order $\alpha, \beta, \gamma, \delta$), the only possible correlated errors that can arise are $X_\beta X_\gamma X_\delta$, $X_\gamma X_\delta$,  or $X_\delta$, corresponding to BBR transitions on data atoms $\beta$, $\gamma$, or $\delta$, respectively. For the stabilizers of Eq.~(\ref{eq:steane-stabilizer}), these errors will all give rise to distinct error syndromes upon measuring $Z^{\otimes 4}$ stabilizers and can thus be corrected (see Appendix \ref{sec:postponing-app}). 
This can substantially reduce the number of measurements required to implement our protocol, making it more feasible for near-term experiments. A similar procedure can be applied to measure the $Z^{\otimes 4}$ stabilizers.

\subsubsection{Logical Operations}

{\bf Logical  Hadamard, Paulis, and $S$ gate. } One particular advantage of the Steane code is the transversality of the logical Hadamard, Pauli, and $S = \textrm{diag}(1,i)$ gates~\cite{Steane96}. Specifically, the logical Hadamard simply consists of a Hadamard on each physical qubit:
\begin{equation}
H_L = \otimes_{j=1}^7 H_j.
\end{equation}
These operations can be performed without ever populating the Rydberg state, and hence without introducing Rydberg gate errors. Similar decompositions exist for the $S$ gate and the Pauli gates $X$, $Y$, and $Z$.

{\bf Logical controlled-phase gate.} 
The controlled-phase gate in the Steane code is also transversal \cite{Steane96}:
\begin{equation}
\label{eq:transversal-CZ}
\CZ_{AB} = \bigotimes_{j_A=j_B=1}^7 \CZ(j_A,j_B).
\end{equation}
We can thus implement a logical controlled-phase operation by performing only seven physical controlled-phase operations and probing for BBR errors in between each physical controlled-phase gate (to convert them to $Z$-type errors). This eliminates the possibility of correlated multi-qubit errors within a single logical qubit.

{\bf Logical Toffoli gate. } 
To implement the Toffoli gate fault-tolerantly and complete our universal gate set, we implement the logical $\CCZ$ gate where the target qubit has been conjugated by Hadamard gates. While this gate is not transversal in the Steane code, it may still be decomposed into a product of physical $\CCZ$ gates in a round-robin fashion~\cite{Yoder16} (see Appendix~\ref{sec:ryd7-blockade-app} for a derivation):
\begin{equation}
\label{eq:ccz-round-robin}
\CCZ_{ABC} = \prod_{j_A,k_B,l_C \in \{ 1,2,3\}} \CCZ(j_A, k_B, l_C),
\end{equation}
so that a logical $\CCZ$ operation can be implemented using 27 physical $\CCZ$ operations. In the Rydberg setup, this is implemented with the three-qubit Rydberg gate $R(j_A, k_B; l_C) = \text{diag}(1,-1,-1,-1,-1,-1,-1,-1)$ and conjugating all involved data qubits by Pauli-$X$. To avoid propagation of correlated errors resulting from an input $X$ error which does not commute with these Rydberg gates, we begin by fault-tolerantly measuring all the $Z^{\otimes 4}$ stabilizers, and correcting any detected errors; it is simple to verify that this protocol can only result in single-qubit $Z$ errors. This can also be achieved in a more resource-efficient manner by requiring that the stabilizer measurements immediately preceding every logical $\CCZ$ gate be done in a way which measures all $Z^{\otimes 4}$ stabilizers last. Furthermore, Rydberg population detection (followed by conversion to $Z$-type errors, if necessary) is performed after every Rydberg gate, but stabilizers do not need to be measured until the very end; this is because only $Z$ errors occur during the gate operations. In this way, the logical $\CCZ$ satisfies the fault-tolerance property.

\begin{figure}[t]
\includegraphics[width=0.95\textwidth]{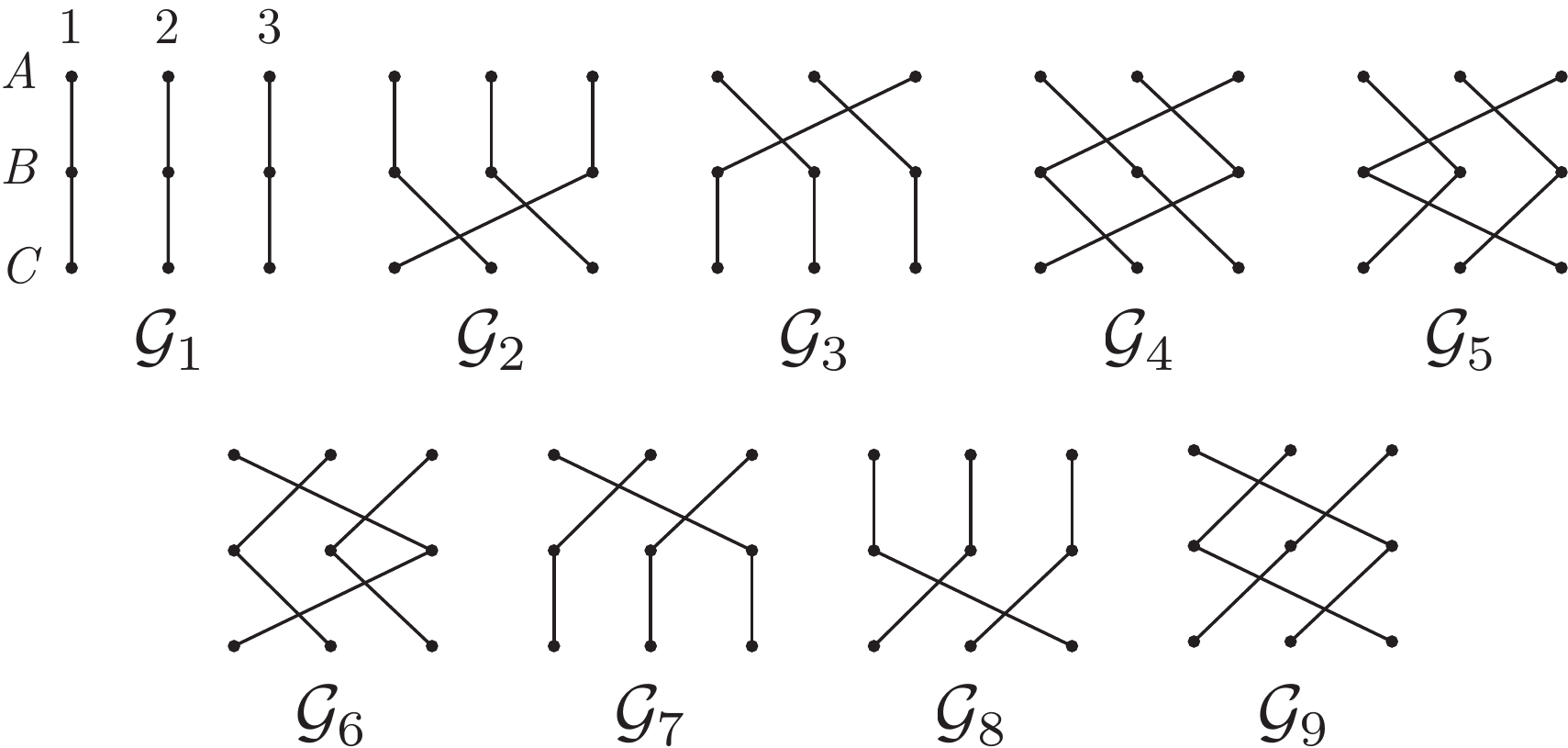}
\caption{Reordering of physical gates in performing the logical $\CCZ$ operation. For each logical qubit, only the first three data qubits are shown, since the other data qubits are not involved in the logical gate. Within each group $\mathcal{G}_i$, the Rydberg gates $R(a,b;c)$ are ordered by increasing index of the physical control qubit $a$ (i.e.\ the data qubit of $A$ involved in the gate).}
\label{fig:ccz-reordering}
\end{figure}

Although the physical implementation of the $\CCZ$ gate is not transversal, the physical gates may be reordered as they all commute with each other. In doing so, we can eliminate some but not all of the intermediate Rydberg population detection steps, to reduce the total number of measurement operations as we did for the fault-tolerant stabilizer measurements. Specifically, we group the three-qubit physical Rydberg gates of the protocol into nine groups of three, $\mathcal{G}_1, ..., \mathcal{G}_9$, so that each physical qubit $j_A, k_B, l_C \in \{ 1,2,3\}$ is used in every group. One example of such a grouping $\mathcal{G}_1, ..., \mathcal{G}_9$ is shown in Figure \ref{fig:ccz-reordering}. With this reordering, detection for Rydberg leakage only needs to be performed after each group $\mathcal{G}_i$. This is because a Rydberg leakage error can only result in the blockading of the last two, the last, or no Rydberg gates within a group $\mathcal{G}_i$, and these cases correspond to disjoint possible sets of stabilizer eigenvalues $(g_2,g_3)$ for the three logical qubits (see Appendix~\ref{sec:postponing-app}).

The Hadamard and $\CCZ$ gates together form a universal gate set for quantum computation \cite{Shi03}, so we have demonstrated a scheme to construct any quantum operation on the code space fault-tolerantly against BBR errors.

\subsubsection{Logical state preparation}

\begin{figure}[t]
\includegraphics[width=0.7\textwidth]{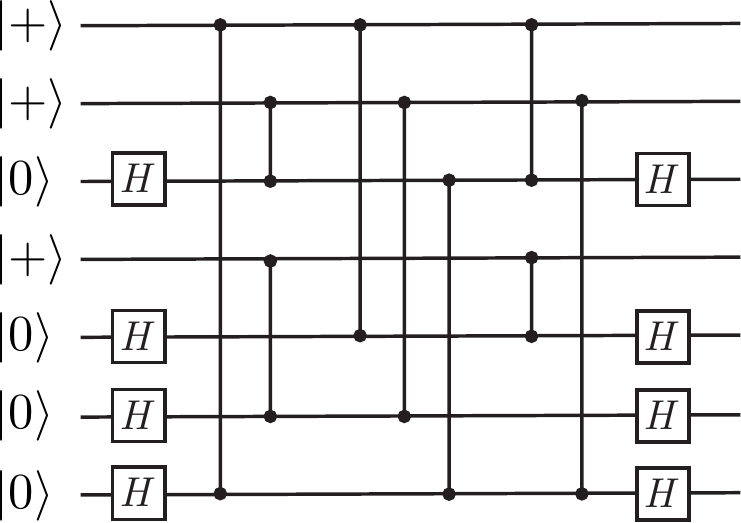}
\caption{Protocol to prepare the logical $\ket{0}_L$ state for the Steane code.}
\label{fig:steane-state-prep}
\end{figure}

Finally, we show that we can prepare the logical $\ket{0}_L$ state in a fault-tolerant manner. The most straightforward preparation of this state uses Steane's Latin rectangle encoding method, whose circuit is shown in Figure~\ref{fig:steane-state-prep} \cite{Steane96}. In the Rydberg setup, we replace controlled-NOT gates by Rydberg controlled-phase gates with target qubit conjugated by Hadamard gates. Because the $Z$ errors associated with Rydberg gates commute with controlled-phase operations, to leading order in $p_{\text{tot}}$, there will be at most one Pauli-$Z$ error among the three data qubits initially in the $\ket{+}$ state, and at most one Pauli-$X$ error among the four data qubits initially in the $\ket{0}$ state. Although this could be a two-qubit error, it is correctable because the Steane code identifies and corrects $X$ and $Z$ errors separately. In this procedure, we have assumed we detect for Rydberg population arising from BBR errors after each physical entangling gate and convert these errors to $Z$ errors as necessary. In this way, by applying one round of stabilizer measurements and error correction, we will obtain (to leading order in $p_{\text{tot}}$) a logical $\ket{0}_L$ state with a Pauli error on at most one physical qubit. 

\subsection{FTQC with BBR and RD errors}
\label{sec:ftqc-lf}

To address RD errors and intermediate state scattering, we must consider two new classes of leakage errors following the discussion of Section \ref{sec:error-models}: (1) leakage into the original Rydberg state $\ket{r}$ and (2) leakage into the other hyperfine ground states, which we will also call ``non-Rydberg leakage.'' The first class of errors is similar to the quantum jumps in the BBR error model, and can be detected and corrected in the same way using an ancilla qubit. In the following sections, we will group this error together with BBR errors and refer to them as ``Rydberg leakage'' errors.

\begin{figure}[t]
\includegraphics[width=\textwidth]{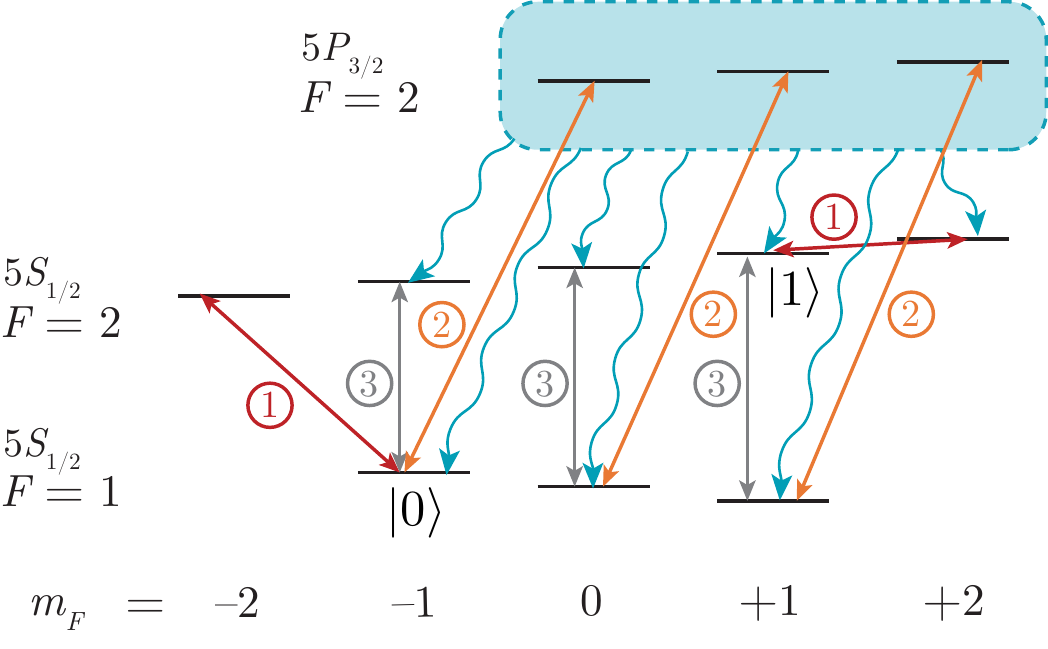}
\caption{Illustration of the optical pumping protocol to convert non-Rydberg leakage errors to Pauli-$Z$ errors in a $^{87}$Rb atom. First, we apply $\pi$ pulses $|1 \rangle \leftrightarrow |F=2, m_F=2 \rangle$ and $|0 \rangle \leftrightarrow |F=2, m_F=-2 \rangle$ (red arrows). In the second step, we use $\sigma^+$ light to excite states in the $F=1$ ground state manifold to the $5P_{3/2}$ $F=2$ manifold (orange arrows). These states decay quickly back into the ground state manifold, as indicated by light blue wavy arrows. Thirdly, we apply resonant $\pi$ pulses  $|F=2,m_F \rangle \leftrightarrow |F=1,m_F \rangle$ (grey arrows). The second and third steps are repeated until all population with $m_F \geq -1$ has been transferred to the stretched state $|F=2, m_F=2 \rangle$. Finally, the first step (red) is repeated to restore the qubit state populations.} 
\label{fig:optical-pumping}
\end{figure}

On the other hand, we demonstrate that leakage to other states in the hyperfine manifold can be converted into Pauli-$Z$ type errors using optical pumping. For example, for $^{87}$Rb, we design the novel optical pumping protocol shown in Figure~\ref{fig:optical-pumping}. One crucial property of this optical pumping procedure is that it does not affect the qubit coherence when there is no error.
Furthermore, notice that while leakage in traditional QEC settings may be particularly difficult to address, requiring additional entangling gates or ancilla qubits, the particular multi-level structure of neutral atoms allows for efficient correction of these errors. Notably, this optical pumping can be performed without the need for qubit measurement and feed-forward corrections, allowing for efficient implementation in experiments. 

The correction of non-Rydberg leakage errors can be incorporated into the fault-tolerant protocols of the previous section by performing this procedure between the Rydberg entangling gates. Thus, our protocols from the previous section will be fault-tolerant against generic intrinsic Rydberg decay errors. Furthermore, note that when considering this full error model including both BBR and RD events, it is no longer necessary to swap population between the $|1 \rangle$ state and the stretched ground state $|F=I+1/2,m_F=I+1/2 \rangle$ when addressing Rydberg leakage errors (i.e., one can omit Steps 1 and 3 in Appendix~\ref{sec:rydberg-leakage-app}); instead, the Rydberg population can be pumped directly to the $|F=I+1/2,m_F=I+1/2 \rangle$ state, converting it into a non-Rydberg leakage error which is corrected by optical pumping. The full protocols for fault-tolerant stabilizer measurement, the logical controlled-phase gate, and the logical $\CCZ$ gate are given in Algorithms \ref{alg:7q-x-stab}-\ref{alg:7q-ccz}.

\begin{algorithm*}[t]
  \caption{Fault-tolerant method to measure $X^{\otimes 4}$ stabilizers for Rydberg 7-qubit code.}
  \label{alg:7q-x-stab}
   \begin{algorithmic}
   \State \begin{enumerate}[label=\arabic*)]
   \item For each $X^{\otimes 4}$ stabilizer $X_\alpha X_\beta X_\gamma X_\delta$:
   \begin{enumerate}[label=\alph*.]
   \item Initialize ancilla qubit $A_2$ to $\ket{+}$ state.
   \item Apply gate $Z_j H_j$ to all data qubits $j \in \{\alpha, \beta, \gamma, \delta\}$.
   \item For each $j \in \{\alpha, \beta, \gamma, \delta\}$, apply the Rydberg gate $R(A_1;D_j)$. If $j=\delta$, use ancilla qubit $A_1$ to detect for Rydberg population as described in Section~\ref{sec:ryd-bbr}; if a Rydberg leakage error is detected, convert it to a non-Rydberg leakage error $|F=I+1/2,m_F=I+1/2 \rangle \langle 1|$. Finally, use the optical pumping technique of Section \ref{sec:ftqc-lf} to convert any possible non-Rydberg leakage error into a possible single-qubit $Z$ error.
    \item Apply Hadamard gates to all data qubits $j \in \{\alpha, \beta, \gamma, \delta\}$.
  \item Measure $A_2$ in the $X$ basis.
  \item If $A_2$ measurement yields $-1$, break.
   	\end{enumerate}
   	\item If any stabilizers are measured to be $-1$: 
   	\begin{enumerate}[label=\alph*.]
   	\item Measure all $X^{\otimes 4}$ stabilizers again, this time in the unprotected way and without checking for leakage. There was either already an error in the input, or an error occurred in the initial measurement process. The resulting outcomes will then be the correct stabilizer values to leading order in $p_{\textrm{tot}}$. 
   	\end{enumerate}
   	\end{enumerate}
   \end{algorithmic}
\end{algorithm*}

\begin{algorithm*}[t]
  \caption{Fault-tolerant logical $\CZ$ for Rydberg 7-qubit code.}
  \label{alg:7q-cz}
   \begin{algorithmic}
   \State \begin{enumerate}[label=\arabic*)]
   \item Apply single-qubit $Z$ gates to all physical control and target qubits.
   \item For each $j = 1, 2, ... , 7$:
   \begin{enumerate}[label=\alph*.]
       \item Apply the two-qubit Rydberg gate $R(C_j; T_j)$.
       \item Use ancilla qubit $A_1$ to detect for Rydberg population as described in Section~\ref{sec:ryd-bbr}.
       \item If a Rydberg leakage error is detected, convert it to a non-Rydberg leakage error $|F=I+1/2,m_F=I+1/2 \rangle \langle 1|$.
   \end{enumerate}
  \item Use the optical pumping technique of Section \ref{sec:ftqc-lf} to convert any possible non-Rydberg leakage error into a possible single-qubit $Z$ error.
   	\end{enumerate}
   \end{algorithmic}
\end{algorithm*}

\begin{algorithm*}[t]
  \caption{Fault-tolerant logical $\CCZ_{ABC}$ for Rydberg 7-qubit code.}
  \label{alg:7q-ccz}
   \begin{algorithmic}
   \State \begin{enumerate}[label=\arabic*)]
   \item Apply $X$ gate to all physical qubits $j_A, k_B, l_C \in \{ 1,2,3\}$.
   \item For each group $\mathcal{G}_i$ of physical three-qubit Rydberg gates to apply (where $\mathcal{G}_i$ are ordered as discussed in the main text or Figure \ref{fig:ccz-reordering}):
   \begin{enumerate}[label=\alph*.]
   \item Apply gates in $\mathcal{G}_i$.
   \item Use ancilla qubit $A_1$ to detect for Rydberg population as discussed in Section \ref{sec:ryd-bbr}. If Rydberg leakage is detected: 
   \begin{enumerate}[label=\roman*)]
\item Convert this leakage error to a possible single-qubit $X$ error.
\item Measure stabilizer eigenvalues $g_2$ and $g_3$ for each logical qubit in an unprotected way. This is safe because an error already occurred.
	\item  Apply the appropriate correction circuit for the correlated error (since the possible correlated errors all result in disjoint sets of possible syndromes; see Appendix~\ref{sec:postponing-app}). 
	\item Measure $Z^{\otimes 4}$ stabilizers for all logical qubits in an unprotected way to detect for a possible single-qubit $X$ error induced by step i) above; correct this error if found.
	\item The remaining three-qubit Rydberg gates needed to implement the logical $\CCZ$ operation can all be applied in an unprotected way.
   \end{enumerate}
   \item  Use the optical pumping technique of Section \ref{sec:ftqc-lf} to convert any possible non-Rydberg leakage error into a possible single-qubit $Z$ error.
   \end{enumerate}
   \item Apply $X$ gate to all physical qubits $j_A, k_B, l_C \in \{ 1,2,3\}$.
   	\end{enumerate}
   \end{algorithmic}
\end{algorithm*}

While the above discussion has focused on intrinsic RD errors, the non-intrinsic errors of Section \ref{sec:other-errors} can also be incorporated into our FTQC protocols. Specifically, the errors resulting from Rydberg laser imperfections such as intensity and phase fluctuations only cause Pauli-$Z$ errors and single-qubit Rydberg leakage errors, so they are already addressed within our current framework. Similarly, atom loss can be detected by using an ancilla qubit and performing a small leakage detection circuit; this is discussed in Appendix~\ref{sec:atom-loss-app}. In this case, if a reservoir of atoms is available, we can also convert the atom loss error into a single-qubit Pauli-$X$ or $Z$ error, for instance by replacing the lost atom with a new atom initialized to the $\ket{0}$ state.

\subsection{Comparison to existing fault-tolerant quantum computing protocols}
\label{sec:comparison}

To demonstrate the significance of our Ryd-7 FTQC protocol and emphasize the importance of considering specific error models when designing QEC approaches, we now compare our model with existing general-purpose FTQC schemes proposed in Refs.~\cite{Chao18,Chao18b,Yoder16}. Specifically, we compare the costs of measuring stabilizers and implementing fault-tolerant logical operations, using as metrics the number of two- and three-qubit entangling operations required for the physical qubits, and the minimum number of ancilla qubits needed. Details on how these numbers can be obtained for the Ryd-7 protocol are provided in Appendix~\ref{sec:resource-cost-app}.

Table \ref{tab:stab-comparison} compares the minimum number of two-qubit gates and ancilla qubits required for fault-tolerant stabilizer measurement (and associated error correction) in various QEC proposals. The results for general-purpose FTQC protocols for the 7- and 15-qubit CSS/Hamming codes are based on the ``flagged syndrome extraction'' procedures presented in Refs.~\cite{Chao18,Yoder17,Chao18b}. For each protocol, we separately present the resource cost for cases without any errors and the worst-case cost when an error is present (numbers in parentheses), as the former case is typically much more probable. While the number of ancilla qubits required is the same for all cases, we find that our protocol requires the smallest number of entangling operations in either case even though we must detect for leakage, an additional kind of error not considered in Refs.~\cite{Chao18,Yoder17,Chao18b}.

Similarly, Table \ref{tab:gate-comparison} demonstrates this comparison for the fault-tolerant logical $\CCZ$ gate, where our improvements are striking. The general-purpose implementation of this non-Clifford gate for three logical qubits in the $7$-qubit Steane code is given by Yoder, Takagi, and Chuang \cite{Yoder16}; while this implementation requires only a modest number of physical two- and three-qubit gates, it requires a considerable overhead of $72$ additional ancilla qubits, making an experimental demonstration very challenging. On the other hand, while Chao and Reichardt's proposal~\cite{Chao18b} for a fault-tolerant Toffoli gate using the $[[15,7,3]]$ code significantly reduces the ancilla qubit count, the number of physical entangling operations is substantial. Our protocol uses only 2 ancilla qubits compared with 72 required in Yoder, Takagi, and Chuang~\cite{Yoder16}, while using significantly fewer entangling operations (e.g.~60 two-qubit gates) than Chao and Reichardt~\cite{Chao18b} (1416 two-qubit gates) even in the unlikely scenario where we must correct for an error. We note that while our protocol does use more three-qubit entangling gates than Ref.~\cite{Yoder16}, such gates are nearly as straightforward to implement as two-qubit $\CZ$ gates in the Rydberg atom setup (see Section~\ref{sec:summary}).

These results clearly demonstrate the advantage of considering a hardware-specific error model and leveraging the unique capabilities of the Rydberg setup when designing FTQC schemes. In particular, even though we must correct for additional errors not considered in traditional settings, we can still dramatically reduce the required number of entangling gates or ancilla qubits.

\subsection{Scalable implementation}
\label{sec:scalability}

We now discuss some more details regarding the scalable implementation of our protocols, including potential geometrical layouts of physical qubits, resource trade-offs, and residual error rates.

{\bf Geometrical considerations. } 
One particular advantage of the Rydberg atom platform is the flexibility in allowing arbitrary geometrical arrangements of atoms. Motivated by recent experimental demonstrations of near-deterministic loading and rearrangement of neutral atoms into regular lattice structures, we propose scalable FTQC architectures in which logical qubits form a coarser lattice on top of the lattice of physical atoms. For the Ryd-7 scheme, one natural layout in a two-dimensional atomic array setup could comprise placing physical atoms on the vertices of a triangular lattice (Figure \ref{fig:overview-figure}a). In this geometry, the hexagonally shaped logical qubits (dark blue dotted hexagons) form a coarser triangular lattice, with ancilla qubits (A, pink) placed on the edges of this coarser lattice to mediate error correction and logical gates. Fault-tolerant universal quantum computation can be performed if nearest-neighbor logical qubits can be entangled; because physical entangling gates can only be implemented between atoms within a blockade radius $R_B$, this defines a minimum required value of $R_B$ in terms of the closest atom-atom separation $d$. Upon examining the physical gates required to implement the logical operations for the seven-qubit code, we find that the requirement in this case is $R_B > 3d$ (dotted grey line), an interaction range which has already been demonstrated in recent experiments~\cite{Bernien17}. Details on the derivation of this minimal blockade radius can be found in Appendices~\ref{sec:blockade-app} and~\ref{sec:ryd7-blockade-app}. This requirement on $R_B$ can be further reduced if atoms can be moved in between certain logical operations while preserving coherence between the hyperfine ground states~\cite{Bluvstein21b}.

{\bf Resource tradeoffs. } 
For any experiment, resource trade-offs may be made to minimize the total logical error probability. For instance, if the timescale of one round of measurements is much larger than typical gate times (as is the case in certain atomic setups), one may wish to reduce the number of measurement shots required at the expense of performing additional operations. This can be incorporated into our protocol by incoherently driving Rydberg states to the low-lying $P$ state after each entangling gate to convert any possible Rydberg leakage error into the non-Rydberg leakage $|F=I+1/2,m_F=I+1/2 \rangle \langle 1|$. In this case, ancilla measurements are no longer necessary to detect and correct for Rydberg leakage errors, but this incoherent pumping would be done after every gate, regardless of whether an error had actually occurred. Alternatively, the number of entangling gates can be further reduced at the cost of additional measurements.

{\bf Improvements. } 
The FTQC protocol presented in this section relies  upon selection rules which impose restrictions on the possible RD error channels. Specifically, as mentioned in Section \ref{sec:error-models}, to leading order in the error probability, we ignored the decay channel $|0 \rangle \langle 1|$ arising from RD. Given the low branching ratio (determined numerically to be $\sim 10^{-3}$ in $^{87}$Rb, see Appendix \ref{sec:transition-rates-app}) from the stretched Rydberg state to $|0 \rangle$, this is already a reasonable assumption; however, several approaches can be taken to suppress the probability of such errors even further. First, this probability can be reduced by a factor of roughly 3 or 4 by employing a ``shelving'' procedure in which population in the $|0 \rangle$ state is swapped with  the stretched ground state $|F=-m_F=I+1/2 \rangle$ before and after each entangling gate, due to the lower branching ratio from $|r \rangle$ to this stretched state. To avoid errors arising from near-degenerate Rydberg transitions in this case, one would also transfer population in the $|1 \rangle$ state to $|F=I-1/2, m_F=1 \rangle$ to perform Rydberg excitation in this case, instead of exciting out of the $F=I+1/2$ manifold. Moreover, by utilizing higher magnetic fields to reduce the branching ratio for RD processes involving large $|\Delta m_F|$, or by using a species with higher nuclear spin (e.g.\ $^{85}$Rb) where the shelving state can be further separated from the stretched Rydberg state, one can suppress the probability of such errors to even higher orders.

\section{Leading-Order Fault-Tolerance with a Repetition Code}
\label{sec:ftqc-repetition}

Given that all Rydberg errors can be converted to the $Z$-type, one may naturally ask whether the full seven-qubit Steane code is even necessary to detect and correct these errors; in particular, one may be tempted to simply use a three-qubit repetition  code in the $X$ basis to detect and correct $Z$-type errors. In such a code, the logical states are 
\begin{equation}
\begin{gathered}
\label{eq:repetition-encoding}
	|+ \rangle_L = |\! +\! ++ \rangle\phantom{,} \\
	|- \rangle_L = |\! -\! -- \rangle,
\end{gathered}
\end{equation}
and stabilizer operators are 
\begin{equation}
\label{eq:repetition-stabilizer}
	g_1 = X_1 X_2, \quad g_2 = X_2 X_3.
\end{equation}
However, direct application of such a repetition code for FTQC is challenging even with this biased noise model, as one must be able to implement every physical gate in the encoding, decoding, stabilizer measurement, and logical gate procedures without introducing Pauli-$X$ or $Y$-type errors at any stage---that is, each gate must be implemented in a {\it bias-preserving} way. 
This requirement can easily be satisfied for certain physical gates such as the Rydberg controlled-phase or collective gates  (after all leakage errors are mapped to Pauli-$Z$ type), but is much more difficult to fulfill for other gates. Specifically, measurement of the stabilizers of Eq.~(\ref{eq:repetition-stabilizer}) requires performing controlled-NOT (CNOT) gates as shown in orange in Figure~\ref{fig:overview-figure}e.

\begin{figure}
\includegraphics[width=0.58\textwidth]{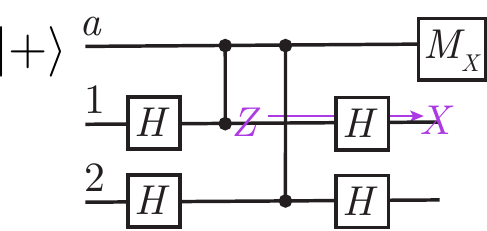}
\caption{Circuit to measure the stabilizer $X_1 X_2$ for the repetition code. CNOT gates must be performed between the ancilla qubit and data qubits $1$ and $2$. A standard implementation of the CNOT gate using Rydberg controlled-phase gates conjugated by single-qubit Hadamard gates on the target qubits would not be bias-preserving, as a $Z$ error on a target qubit during a controlled-phase gate would become an $X$ error once the final Hadamard gate is applied (purple).}
\label{fig:repetition-stabilizer-measurement}
\end{figure}

While a standard implementation of the CNOT gate in a Rydberg setup would comprise the Rydberg controlled-phase gate conjugated by single-qubit Hadamard gates on the target qubit, this would not be bias-preserving: for example, a $Z$ error on a target qubit during a controlled-phase gate would become an $X$ error once the final Hadamard gate is applied (Figure~\ref{fig:repetition-stabilizer-measurement}). 
In other setups, where a $\pi$-rotation of the target qubit about the $\hat{x}$ axis on the Bloch sphere can be performed conditioned on the state of the control qubit (e.g.\ by engineering a $H_{\textrm{int}} = ZX$ interaction), an over-rotation or under-rotation error would also translate to an $X$ error and violate the bias-preserving constraint.

Indeed, the implementation of a bias-preserving CNOT may seem unfeasible at first, in light of a no-go theorem proven in Ref.~\cite{Guillaud19}: a bias-preserving CNOT gate is not possible between two qubits encoded in systems where the underlying Hilbert space is finite-dimensional, because the identity gate cannot be smoothly connected to CNOT while staying within the manifold of bias-preserving operations. One way to circumvent this no-go theorem was recently developed for circuit QED systems in Refs.~\cite{Guillaud19,Puri19}, where the qubits can be encoded in the continuous phase space of the photon field, and the dominant source of error---photon loss---can be manipulated via parametric driving schemes to cause only $Z$-type errors. In our setup, we circumvent the no-go theorem using the special fact that certain pulses in our finite-dimensional atomic system---the pulses between hyperfine states---can be implemented at very high fidelities, so that our leading-order errors arise only from Rydberg pulse imperfections and Rydberg state decay. This allows us to develop a  novel laser pulse sequence for entangling Rydberg atoms that directly implements a CNOT or Toffoli gate while preserving the noise bias. 
Our protocol can be applied on any atomic species with sufficiently high nuclear spin ($I \geq 5/2$). For concreteness, we will illustrate the protocol using the example case of $^{85}$Rb throughout the section.

\subsection{Bias-preserving CNOT in a Rydberg atom setup}
\label{sec:rydberg-bias-preserving}

\begin{figure*}
\includegraphics[width=0.75\textwidth]{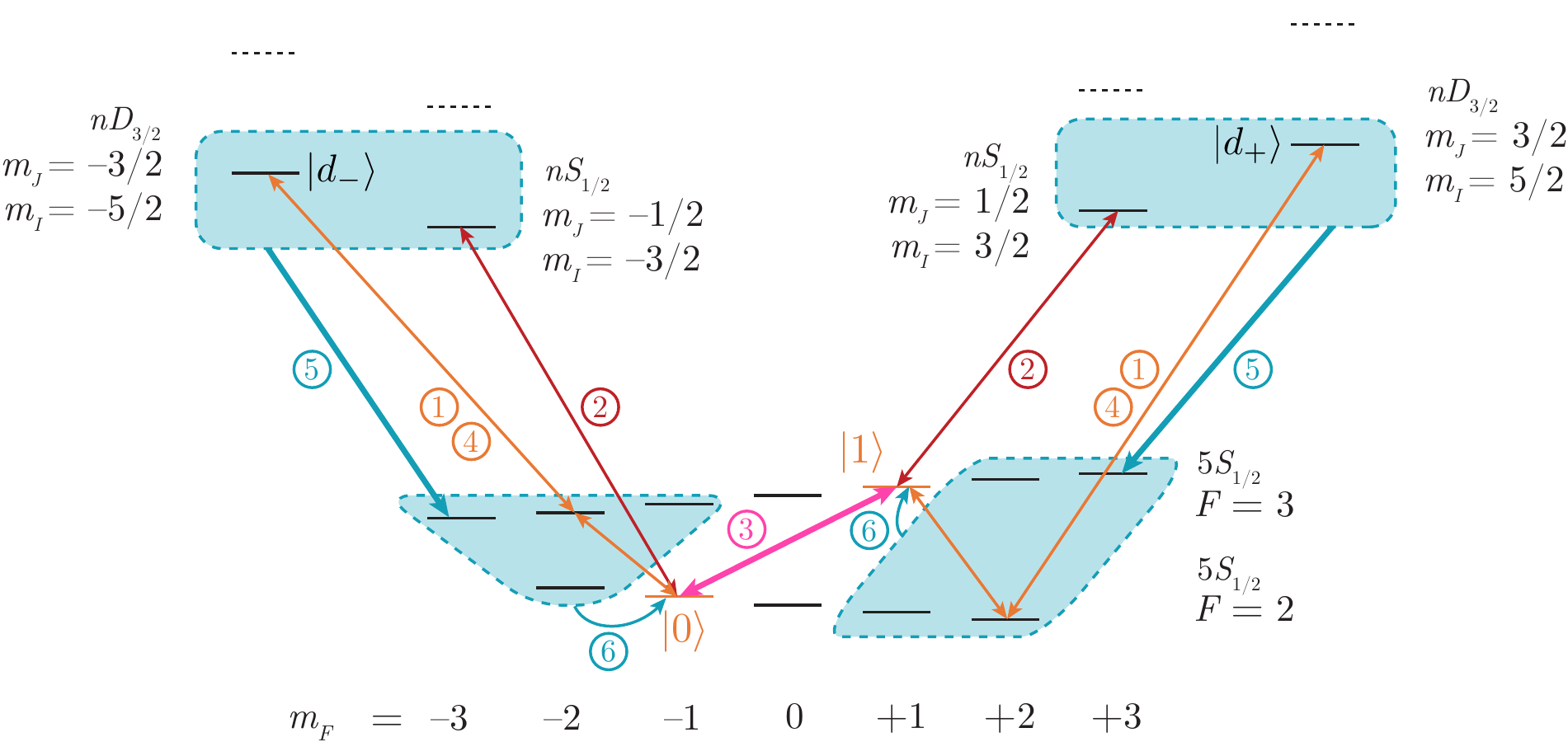}
\caption{Pulse sequence for the target atom in a bias-preserving CNOT gate between $^{85}$Rb atoms. Rydberg pulses are resonant if and only if no nearby Rydberg population is present; otherwise, the Rydberg levels are shifted due to the blockade effect (dotted levels). This pulse sequence eliminates target atom $X$ errors in the standard implementation of CNOT shown in Figure~\ref{fig:repetition-stabilizer-measurement}. Step 1: Coherent transfer of population from the qubit states to stretched Rydberg states $|d_\pm \rangle \equiv |nD_{3/2}, m_J = 3/2, m_I = I =5/2\rangle$. To do this, we first apply hyperfine $\pi$ pulses $|1\rangle \leftrightarrow |F=2,m_F=2\rangle$ and $|0\rangle \leftrightarrow |F=3,m_F=-2\rangle$, then apply Rydberg $\pi$ pulses $|F=2,m_F=2\rangle \leftrightarrow |d_+\rangle$, $|F=3,m_F=-2\rangle \leftrightarrow |d_-\rangle$, and finally reapply the hyperfine $\pi$ pulses $|1\rangle \leftrightarrow |F=2,m_F=2\rangle$ and $|0\rangle \leftrightarrow |F=3,m_F=-2\rangle$ (orange arrows). 
 Step 2: Apply resonant $\pi$ pulses
from the qubit states to the Rydberg states $|1 \rangle \leftrightarrow  |nS_{1/2}, m_J =  1/2, m_I =  3/2\rangle$ and $|0 \rangle \leftrightarrow |nS_{1/2}, m_J =  -1/2, m_I = -3/2\rangle$ (red arrows). Step 3: Apply a resonant $\pi$ pulse between the $|0 \rangle$ and $|1 \rangle$ ground states (thick pink arrow). Step 4: Repeat Step 1, but use $-\pi$ instead of $\pi$ pulses on all transitions. Step 5: Incoherently drive any remaining Rydberg population into stretched ground states (thick blue arrows). Specifically, send Rydberg states with $m_J+m_I > 0$ (respectively, $<0$) to a stretched $5P$ state with $F = m_F = I+3/2$ ($F=-m_F=I+3/2$), which decays quickly and only to the stretched ground state with $F=m_F=I+1/2$ ($F=-m_F=I+1/2$). Step 6: Use optical pumping techniques (see Appendix~\ref{sec:optical-pumping-app} for details) to map states outside the computational subspace with $m_F > 0$ (respectively, $m_F < 0$) to the qubit state $|1 \rangle$ ($|0 \rangle$) (thin blue arrows). 
}
\label{fig:bias-preserving-cnot-target}
\end{figure*}

As shown in Figure~\ref{fig:repetition-stabilizer-measurement}, the standard implementation of a CNOT gate in a Rydberg system is not bias-preserving. In particular, given the error model for Rydberg gates, $X$ errors on the target qubit can be induced in two ways.
First, the target qubit could directly undergo a Rydberg error (e.g.\ radiative decay) during the controlled-phase gate, resulting in a Pauli-$Z$ error that is transformed into an $X$ error after the Hadamard gate (purple in Figure~\ref{fig:repetition-stabilizer-measurement}).
Alternatively, the control atom could decay from the Rydberg state to the ground state at some point during the controlled-phase gate, so that the target qubit Rydberg pulses, which should have been blockaded, are now resonant during the controlled-phase gate. This results in a two-qubit correlated error between the control and target atoms, where the target atom undergoes an $X$-type error. 

Here, we begin by introducing a novel entangling gate pulse sequence for Rydberg atoms to address the target atom $X$ errors. In this discussion, we first assume that the Rydberg pulses on the target atom are either all resonant or all blockaded; that is, we ignore the possibility of a neighboring Rydberg atom decaying during the target atom sequence. We then  include this effect and also eliminate the correlated errors by introducing an ancilla qubit and making use of two Rydberg states with different blockade radii.

To remove the target atom $X$ errors, we wish to design an entangling gate protocol which uses Rydberg states to conditionally swap $|0 \rangle$ and $|1 \rangle$ population directly, without the change-of-basis from Hadamard gates. This can be accomplished for atomic species with high enough nuclear spin ($I \geq 5/2$). 
We consider qubits encoded in the $^{85}$Rb clock states $|1 \rangle \equiv |F = I+1/2, m_F = +1 \rangle$, $|0 \rangle \equiv |F = I-1/2, m_F = -1 \rangle$ (orange levels in Figure \ref{fig:bias-preserving-cnot-target}), which have a magnetic field-insensitive transition frequency at low fields. The protocol then proceeds as illustrated in Figure \ref{fig:bias-preserving-cnot-target}.

The first step of the procedure aims to transfer population in the qubit state $|1\rangle$ (respectively, $|0\rangle$ to the Rydberg state $|d_+ \rangle \equiv |nD_{3/2}, m_J = 3/2, m_I = I =5/2\rangle$ (respectively, $|d_- \rangle \equiv |nD_{-3/2}, m_J = -1/2, m_I = -I =-5/2\rangle$) conditionally, dependent on the state of a control atom. This is achieved because the Rydberg pulses from the qubit states to $|d_\pm \rangle$ are resonant if and only if there are no neighboring atoms in $|r_\pm \rangle$ or nearby Rydberg states. Since each stretched Rydberg state predominantly decays only into ground states with $|\Delta m_F| = |\Delta(m_J+m_I)| \leq 2$ during RD processes, the $|0 \rangle$ and $|1 \rangle$ populations will not be mixed by Rydberg state decay; however, due to the possible decay channels $|F=2,m_F=2 \rangle \langle d_+|$ and $|F=3,m_F=-2\rangle \langle d_-|$, it is possible that the first step fails to excite the atom into a Rydberg state even in the absence of nearby Rydberg population.  Consequently, in the second step, we again attempt to transfer the qubit states to Rydberg states, this time using resonant $\pi$ pulses $|1 \rangle \leftrightarrow  |nS_{1/2}, m_J =  1/2, m_I =  3/2\rangle$ and $|0 \rangle \leftrightarrow |nS_{1/2}, m_J =  -1/2, m_I = -3/2\rangle$. 
Then, in the third step, the population in the qubit states is swapped via the $\pi$ pulse $|0 \rangle \leftrightarrow |1 \rangle$. We note that this only swaps population if nearby Rydberg atoms prevented transfer out of the qubit manifold in Steps 1 and 2. Step 4 then acts to invert the first step. 

After Step 4, we find that if no Rydberg errors have occurred, the atomic state is restored to the original qubit state (identity map) when no nearby Rydberg population is present, or to the opposite qubit state $|0 \rangle \leftrightarrow |1 \rangle$ otherwise. Rydberg errors can occur only if the pulses of Step 1 are resonant (i.e.\ if no nearby Rydberg atoms are present); moreover, because transitions from $|d_+ \rangle$ (respectively, $|d_-\rangle$) only result in states with $m_F > 0$ ($m_F < 0$), any Pauli errors must be of $Z$-type (for example, projectors $|0 \rangle \langle 0|$, $|1 \rangle \langle 1|$), and any leakage error must be of the form $|m_F > 0 \rangle  \langle 1 |$ or $|m_F < 0 \rangle  \langle 0 |$. 
One can then verify that after the pumping steps (5 and 6), the resulting state is the same as in the error-free case, up to a local error of $Z$ type (e.g.\ $|0 \rangle \langle 0|$, $|1 \rangle \langle 1|$). As before, the error channels for intermediate state scattering and other Rydberg pulse imperfections can be captured by our error model which contains BBR and RD errors.

\begin{figure}
\includegraphics[width=0.8\textwidth]{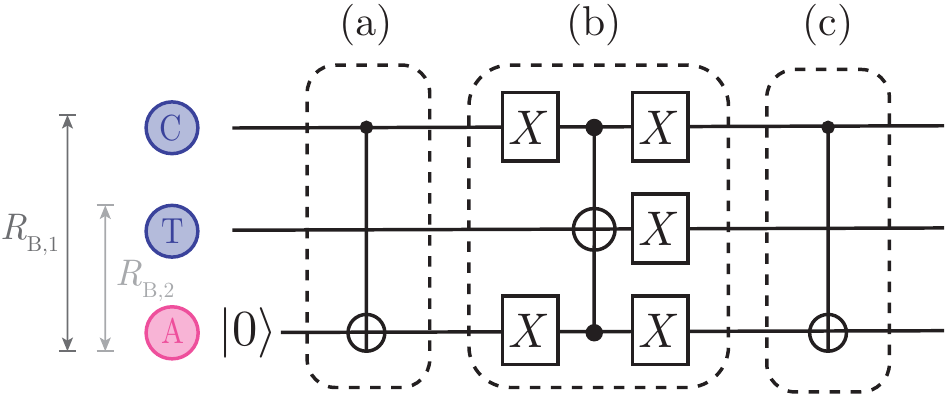}
\caption{Using an ancilla qubit and multiple Rydberg states to eliminate $X$-type errors arising from control qubit decay. The atoms are positioned on a line, such that atom $T$ is in the middle, and the distance between neighboring atoms is $d \equiv d_{{CT}} = d_{{AT}}$. The ancilla qubit is initially prepared in the $|0 \rangle$ state. The protocol consists of three steps, labelled (a)-(c), and can be visualized as a quantum circuit. We use two different pairs of Rydberg $S$ states, $|r_{1,\pm} \rangle$ and $|r_{2,\pm} \rangle$, with blockade radii $R_{B,1}$ and $R_{B,2}$, respectively, such that $R_{B,1} > 2 d$ and $d < R_{B,2} < 2d$. 
Steps (a), (c): Apply a CNOT gate with $C$ as control and $A$ as target. This is done by applying a $\pi$ pulse $|1 \rangle \leftrightarrow |r_{1,+ }\rangle$ on atom $C$, performing the pulse sequence of Figure \ref{fig:bias-preserving-cnot-target} on atom $A$, and applying a $-\pi$ pulse $|1 \rangle \leftrightarrow |r_{1,+ }\rangle$ on atom $C$, so that the Rydberg pulses on $A$ are resonant only if $C$ is not in $|r_{1,+ }\rangle$  (or a nearby Rydberg state). For these steps, the Rydberg levels $|r_\pm \rangle$ in Figure~\ref{fig:bias-preserving-cnot-target} are chosen to be $|r_{1,\pm} \rangle$ (see Table~\ref{tab:bias-preserving-cnot-control}).
Step (b): Apply a three-atom gate between $C$, $A$, and $T$. This is done by applying $\pi$ pulses $|1 \rangle \leftrightarrow |r_{2,+}\rangle$ on both atom $C$ and atom $A$, performing the pulse sequence of Figure~\ref{fig:bias-preserving-cnot-target} on atom $T$, and applying $-\pi$ pulses $|1 \rangle \leftrightarrow |r_{2,+ }\rangle$ on both atom $C$ and atom $A$, so that the Rydberg pulses on $T$ are resonant only if neither $C$ nor $A$ is in $|r_{2,+ }\rangle$  (or a nearby Rydberg state). For this step, the Rydberg levels $|r_\pm \rangle$ in Figure~\ref{fig:bias-preserving-cnot-target} are chosen to be $|r_{2,\pm} \rangle$ (see Table~\ref{tab:bias-preserving-cnot-control}).}
\label{fig:bias-preserving-cnot-control}
\end{figure}

\begin{table}[t]\caption{Rydberg transitions used to implement the bias-preserving CNOT gate between two atoms $C$ and $T$ as shown in Figure~\ref{fig:bias-preserving-cnot-control}.  Within each step, one Rydberg transition ($|1\rangle \leftrightarrow |r_{1,+}\rangle$ or $|1\rangle \leftrightarrow |r_{2,+}\rangle$) is addressed for each ``control'' atom, while two Rydberg transitions ($|0\rangle \leftrightarrow |r_{1,-}\rangle$, $|1\rangle \leftrightarrow |r_{1,+}\rangle$ or $|0\rangle \leftrightarrow |r_{2,-}\rangle$, $|1\rangle \leftrightarrow |r_{2,+}\rangle$) are addressed for each ``target'' atom. $|r_{1,\pm} \rangle$ and $|r_{2,\pm} \rangle$ have different blockade radii $R_{B,1}$ and $R_{B,2}$ as explained in the main text and in the caption of Figure~\ref{fig:bias-preserving-cnot-control}.}\label{tab:bias-preserving-cnot-control}
{\setlength{\tabcolsep}{0.8em}
\begin{tabular}{?c?c|c|c?}
\Xhline{2\arrayrulewidth} & \multicolumn{3}{c?}{{\scshape{Rydberg Transitions Addressed}}} \\ \cline{2-4}
{{\scshape{Step}}}  & {\bf Atom} $C$ & {\bf Atom} $T$ & {\bf Atom} $A$\\ \Xhline{2\arrayrulewidth}
{\bf (a), (c)} & $|1 \rangle \leftrightarrow |r_{1,+} \rangle$  & none & 
\begin{tabular}{@{}c@{}}$|0 \rangle \leftrightarrow |r_{1,-}\rangle$ \\ $|1 \rangle \leftrightarrow |r_{1,+}\rangle$\end{tabular}
\\ \hline
{\bf (b)} & $|1 \rangle \leftrightarrow |r_{2,+} \rangle$  & \begin{tabular}{@{}c@{}}$|0 \rangle \leftrightarrow |r_{2,-}\rangle$ \\ $|1 \rangle \leftrightarrow |r_{2,+}\rangle$\end{tabular} &  $|1 \rangle \leftrightarrow |r_{2,+}\rangle$\\ \Xhline{2\arrayrulewidth}
\end{tabular}}
\end{table}

Having eliminated $X$ errors arising from target qubit Rydberg errors, we now proceed to address the second type of potential $X$ error arising from control qubit decay. The crux here is to utilize multiple Rydberg atoms (e.g.\ a control atom and an ancilla atom) to blockade the target atom if the control is in the $|1 \rangle$ state; in this way, if one of the atoms decays, the remaining Rydberg atom(s) can still ensure (to leading order in the total error probability) that the Rydberg pulses on the target atom do not become resonant. For the simplest case, the bias-preserving CNOT gate can be implemented with one ancilla qubit. Let us assume that the control ($C$), target ($T$), and ancilla ($A$) atoms are placed evenly along a line, with the target atom in between the control and ancilla atoms; the ancilla atom is initialized in the state $|0 \rangle$. We can make use of two sets of Rydberg states, $|r_{1,\pm} \rangle$ and $|r_{2,\pm} \rangle$,  with blockade radii $R_{B,1}$ and $R_{B,2}$, respectively, such that $R_{B,1} > 2  d$ and $d<R_{B,2}<2d$, where $d$ is the distance between neighboring atoms (i.e.\ between $C$ and $T$ or $T$ and $A$); as such, atoms $C$ and $A$ are within the blockade radius $R_{B,1}$, but beyond $R_{B,2}$, whereas neighboring atoms are within the blockade radius $R_{B,2}$. The full bias-preserving CNOT gate between the control and target atoms then consists of the three-step procedure illustrated in Figure~\ref{fig:bias-preserving-cnot-control}, followed by correction of Rydberg leakage errors (as discussed in Appendix~\ref{sec:rydberg-leakage-app}) and optical pumping to eliminate non-Rydberg leakage errors (Figure~\ref{fig:optical-pumping}). 
The Rydberg transitions addressed in each step of Figure~\ref{fig:bias-preserving-cnot-control} are listed in Table~\ref{tab:bias-preserving-cnot-control}.

This protocol is robust against control atom decay errors, as the Rydberg pulses on atom $T$ are resonant only if neither $C$ nor $A$ is excited to the Rydberg state, and one can see that, to leading order in the total error probability, this can only occur if $C$ starts in the $|0 \rangle$ state: first, if $C$ begins in the $|0 \rangle$ state, $A$ must also remain in $|0 \rangle$, so the state of $T$ will not be flipped. On the other hand, if $C$ begins in the $|1 \rangle$ state and no decay events occur during Step (a), $|C,A \rangle = |1,1 \rangle$ after this step. The Rydberg pulses for $T$ are blockaded in Step (b), so its state will be flipped. Finally, if $C$ begins in the $|1 \rangle$ state but decays during the first step, $|C,A \rangle = |1,1 \rangle$ or $|1,0 \rangle$ after this step. The Rydberg pulses for $T$ are still blockaded in Step (b), so its state will be flipped. Finally, Rydberg decay errors in Step (c) will result in projections of the form $|0\rangle \langle 0|$ or $|1 \rangle \langle 1|$, which can be expressed in terms of $Z$ errors. 
In this way, we have eliminated any possible source of $X$ errors arising from the CNOT gate, to leading order in the total error probability. The protocol can also be generalized to implement a bias-preserving Toffoli gate (see  Appendix~\ref{sec:toffoli-app} and Figure~\ref{fig:bias-preserving-toffoli}). Potential improvements leading to suppression at higher orders are discussed in Section~\ref{sec:three-atom-geometry}.

The ability to couple atoms to two sets of Rydberg states $|r_{1,\pm}\rangle$ and $|r_{2,\pm}\rangle$ in our bias-preserving CNOT implementation allows atom $C$ to interact with atom $A$ during Steps (a) and (c) of Figure~\ref{fig:bias-preserving-cnot-control}, but not during Step (b). Alternatively, this tunability of interaction could be achieved with only a single set of addressable Rydberg states $|r_{1,\pm}\rangle$ if the atoms can be rearranged while preserving coherence between hyperfine ground states~\cite{Bluvstein21b}. In this case, one could move atoms in between Steps (a) and (b) to further separate $C$, $T$, and $A$ from each other such that the distance between $C$ and $A$ becomes greater than $R_{B,1}$, while the distance between either of them and atom $T$ remains less than $R_{B,1}$. The atoms can then be returned to their original configuration after Step (b) to allow for interaction between $C$ and $A$ during Step (c).

\subsection{Leading-order fault-tolerance with the repetition code}
\label{sec:ftqc-three-atom}

The bias-preserving operations discussed above allow for a direct implementation of each component of the three-atom repetition code to perform quantum computation with leading-order fault-tolerance on a Rydberg setup. In particular, logical states (\ref{eq:repetition-encoding}) can be prepared or measured fault-tolerantly in the $X$ basis by transversally preparing or measuring each atom. The measurement of stabilizers (\ref{eq:repetition-stabilizer}) can be achieved using the circuit of Figure~\ref{fig:overview-figure}e, where each controlled-NOT gate is done in the bias-preserving way described above; for robustness against errors occurring during this circuit, one must repeat the stabilizer measurement if either $g_1$ or $g_2$ is measured to be $-1$.

A universal set of logical operations can be achieved by implementing a logical Toffoli gate and a logical Hadamard gate as in the seven-qubit case, using the bias-preserving pulse sequences presented above. While not strictly necessary, we will also discuss the implementation of logical controlled-phase and $\CCZ$ gates. These may be of use for simplifying the implementation of certain quantum algorithms, as they do not require the new bias-preserving pulse sequences and can be implemented using the standard method for performing Rydberg-mediated entangling gates as described in Figure~\ref{fig:collective-gates}.

\begin{figure}
\includegraphics[width=0.9\textwidth]{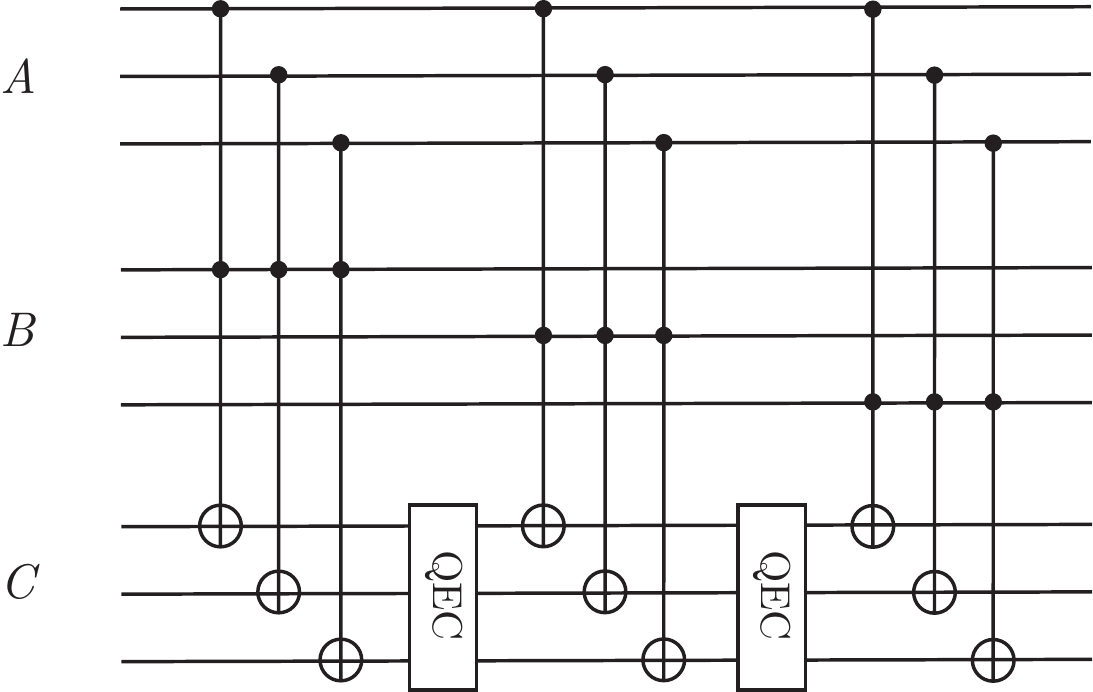}
\caption{Pieceable fault-tolerant implementation of the Toffoli gate in the repetition code.}
\label{fig:repetition-toffoli}
\end{figure}

{\bf Logical Toffoli gate. } 
One important feature of the encoding (\ref{eq:repetition-encoding}) is that the logical $\ket{0}_L$ (respectively, $\ket{1}_L$) state consists of an equal superposition of states with an even (odd) number of physical qubits in the $\ket{1}$ state:
\begin{equation}
\begin{gathered}
|0\rangle_L = \frac{1}{2} \left( |000\rangle + |110 \rangle + |101\rangle + |011\rangle 
\right)\phantom{.} \\
|1\rangle_L = \frac{1}{2} \left(
|111\rangle +
|001\rangle + |010\rangle + |100\rangle  
\right).
\end{gathered}
\end{equation}
From this observation, one can see that the Toffoli gate  $\textrm{CCX}_{ABC}$ with logical control qubits $A$, $B$ and logical target qubit $C$ can be implemented as a product of nine physical Toffoli gates:
\begin{equation}
\textrm{CCX}_{ABC} = \prod_{\substack{j_A,k_B \in \{1,2,3\} \\ l_C = j_A}} \textrm{CCX}(j_A,k_B, l_C).
\end{equation}
Each physical Toffoli gate can be implemented in a bias-preserving fashion as described previously, resulting in at most one physical $Z$ error in each logical qubit, assuming that Rydberg and non-Rydberg leakage errors are converted to possible $Z$ errors after each physical gate. In this case, however, while $Z$ errors on the control qubits $A$ or $B$ would commute with remaining Toffoli gates, a $Z$ error on one of the physical qubits of $C$ could spread to multiple $Z$ errors within $A$ or $B$ after subsequent Toffoli gates if uncorrected. To address this, we order the physical gates as shown in Figure \ref{fig:repetition-toffoli} and perform error correction after every three physical Toffoli operations by measuring the stabilizers (\ref{eq:repetition-stabilizer}); this follows  the {\it pieceable fault-tolerant} implementations of non-transversal gates discussed in Refs.~\cite{Yoder16,Guillaud19}. In this way, after the entire logical gate, there will be at most one physical qubit $Z$ error per involved logical qubit.

\begin{figure}
\includegraphics[width=0.65\textwidth]{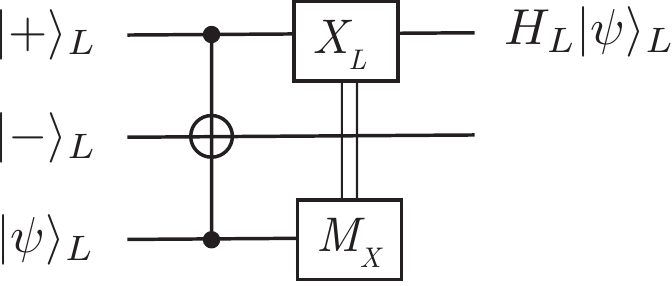}
\caption{Implementing the logical Hadamard in the repetition code using the logical Toffoli gate, as discussed in Ref.~\cite{Guillaud19}.}
\label{fig:repetition-hadamard}
\end{figure}

{\bf Logical Hadamard gate. } 
Unlike the Steane code, the repetition code is not a CSS code, and its logical Hadamard gate is not transversal. However, as discussed in Ref.~\cite{Guillaud19}, the logical Hadamard gate can be implemented using a logical Toffoli gate combined with fault-tolerant measurements in the $X$ basis, as shown in Figure \ref{fig:repetition-hadamard}. The logical Hadamard gate combined with the logical Toffoli or $\CCZ$ gate form a universal set of logical operations.

{\bf Logical controlled-phase gate. } 
A logical controlled-phase operation in the three-qubit code can be implemented using the standard Rydberg pulse sequences for controlled-phase gates between each pair $(j_A, k_B)$ of physical qubits, where $j_A$  and $k_B$ belong  to the encoding of logical qubits $A$ and $B$, respectively:
\begin{equation}
\CZ_{AB} = \prod_{j_A,k_B \in \{1,2,3\}}  \CZ(j_A,k_B).
\end{equation}
To correct for the errors that occur during gates, one should remove any  Rydberg population and apply the optical pumping scheme to convert non-Rydberg leakage errors into possible $Z$ errors after each physical controlled-phase operation. The stabilizers only need to be measured after the entire logical operation, since Rydberg gates can only produce $Z$ errors which commute with all the physical $\CZ$ gates being performed (and hence do not spread to higher-weight errors).

{\bf Logical $\CCZ$ gate. } 
Similarly, a logical controlled-controlled-$Z$ operation between logical qubits $A$, $B$, $C$,
\begin{equation}
\CCZ_{ABC} = \mathbf{1}_A\mathbf{1}_B-\frac{1}{4}(Z_A- \mathbf{1}_A)(Z_B- \mathbf{1}_B)(Z_C- \mathbf{1}_C),
\end{equation}
can be implemented as a sequence of physical $\CCZ$ operations:
\begin{equation}
\CCZ_{ABC} = \prod_{j_A,k_B,l_C \in \{1,2,3\}}  \CCZ(j_A,k_B,l_C).
\end{equation}
As with the case of logical $\CZ$, Rydberg and non-Rydberg leakage errors should be converted to possible $Z$ errors after each physical gate. Notice that even though the logical $\CCZ$ is not transversal, this implementation is leading-order fault-tolerant because any given physical gate can result in at most one physical qubit $Z$ error per logical qubit; since $Z$ errors commute with the remaining gates applied, they do not propagate to become multi-qubit errors. While the $\CCZ$ gate is not strictly needed for the universal gate set given a leading-order fault-tolerant implementation of the logical Toffoli gate, it requires fewer resources to implement than the logical Toffoli as it uses the standard, simpler Rydberg gates $R(C_1,C_2;T)$ instead of the more complicated bias-preserving CNOT pulse sequences (see Table \ref{tab:gate-comparison}). Thus, this operation may be useful for reducing the resource cost of certain quantum algorithms.

\subsection{Scalable implementation}
\label{sec:three-atom-geometry}

We now discuss some important considerations for the scalable implementation of our Ryd-3 protocol, including the geometrical layout, resource requirements, and potential improvements.

{\bf Geometrical layout. } 
Based on the implementations of logical gates, stabilizer measurement, and the underlying bias-preserving CNOT given in the previous sections, we find that a convenient geometry is to place data and ancilla atoms on the vertices of a triangular lattice as shown in Figure \ref{fig:overview-figure}d, with three data atoms comprising a logical qubit. In this configuration, the logical qubits form a coarser triangular lattice, as in the case of Ryd-7. As discussed in Section~\ref{sec:rydberg-bias-preserving}, two Rydberg states with different blockade radii $R_{B,1} > R_{B,2}$ are required to implement the bias-preserving CNOT gate. Based on the interaction ranges required for performing fault-tolerant stabilizer measurements and logical operations as described previously, we find that the larger blockade radius must be greater than $3d$ (dark grey in Figure \ref{fig:overview-figure}d), where $d$ is the nearest-neighbor spacing on the square lattice; this is required for some of the physical gates in the logical $\CCZ$ and Toffoli gates. On the other hand, the smaller blockade radius $R_{B,2}$ should be strictly between $d$ and $2d$ for efficient implementation of the bias-preserving CNOT and fault-tolerant stabilizer measurements (light grey in Figure \ref{fig:overview-figure}d). Details on how to obtain the requirement $R_{B,1} > 3d$ can be found in Appendix~\ref{sec:blockade-app}. 

Alternatively, the data and ancilla atoms can be placed on the vertices of a square lattice in an alternating fashion (see Appendix \ref{sec:ryd3-sq-app}). In this case, the blockade radius requirements are $R_{B,1} > 3.61 d$ and $d < R_{B,2} < 2 d$. For both the triangular lattice and square lattice geometries, experimental developments allowing for rearrangement of atoms while preserving the coherence of hyperfine ground states could be used to further reduce the requirement on $R_{B,1}$ and eliminate the need for a second set of Rydberg states with blockade radius $R_{B,2}$.

{\bf Resource comparison. } We now compare the resource cost of our Ryd-3 protocol with our Ryd-7 approach and the traditional general-purpose proposals of Refs.~\cite{Yoder16,Chao18b}. Compared to the seven-qubit approaches, we find that the number of entangling gates required for extraction of all stabilizers for error correction is significantly reduced due to the smaller number of data atoms and stabilizers per logical qubit, without a substantial increase in the number of required ancillas (Table \ref{tab:stab-comparison}). On the other hand, while the cost of performing a logical $\CCZ$ gate is essentially the same as in Ryd-7, the number of gates required for a logical Hadamard is  larger  (Table \ref{tab:gate-comparison}) because the Hadamard gate is not transversal using the repetition code. Notice that each CNOT gate in a stabilizer measurement translates to two two-atom entangling gates and one three-atom entangling gate in the bias-preserving implementation; this is reflected in Tables \ref{tab:stab-comparison} and \ref{tab:gate-comparison} (more details on obtaining the Ryd-3 resource costs can be found in Appendix~\ref{sec:resource-cost-app}). 
Nevertheless, the number of required gates is still very modest compared to logical operations in other universal FTQC gate sets. As a result, we believe that the substantial resource cost reduction for stabilizer measurements and the improved efficiency in using fewer atoms make the three-atom approach very promising for near-term implementation.

{\bf Improvements. } While our bias-preserving CNOT suppresses $X$-type errors to leading order, the amount of bias preservation is ultimately limited by the decay rate of the stretched Rydberg $D$ state into the qubit states. To further suppress these errors, one can shelve to stretched Rydberg states with higher angular momentum, which would have lower decay rate to the qubit states. Alternatively, one can also use an atomic species with higher nuclear spin, where the qubit states can be separated from the stretched Rydberg state by a larger amount $|\Delta m_F|$. Likewise, one could also increase the magnetic field in the experimental setup to suppress the rate of transitions with high $|\Delta m_F|$. 
To achieve suppression beyond the leading order, one can then use more Rydberg ``shelving'' states in the target atom pulse sequence of Figure \ref{fig:bias-preserving-cnot-target} and more ancillas to suppress the effects of control atom decay.

The Ryd-3 hardware-tailored FTQC approach inherently addresses errors due to Rydberg pulse imperfections in addition to those arising from the finite Rydberg state lifetime, as these errors fall within a subset of the radiative decay errors. As in the Ryd-7 case, the Ryd-3 approach can also be enhanced to further protect against atom loss errors at the expense of additional physical operations by incorporating the atom loss detection scheme described in Appendix~\ref{sec:atom-loss-app} in between Rydberg operations.

\section{Further Considerations Towards Experimental Implementation}
\label{sec:experiments}

In this section, we discuss further considerations on how our FTQC protocols can be implemented in near-term experiments. Recent experiments using neutral alkali atom systems have already achieved near-deterministic trapping, loading, and rearrangement of tens to hundreds of atoms into two-dimensional lattice structures such as the triangular lattice needed for our protocol~\cite{Bluvstein21,Ebadi20,Barredo16,Kim18}. Furthermore, high-fidelity manipulations within the ground state manifold and two- and three-atom Rydberg blockade-mediated entangling gates have been demonstrated \cite{Levine18,Levine19,Graham19}. Blockade interactions between Rydberg atoms separated by three times the lattice spacing, which is the interaction range required for both of our protocols, have also been realized~\cite{Bernien17}.

\subsection{Measurements and feed-forward corrections} 

To perform QEC, an important ingredient is the ability to measure the states of ancilla qubits and/or detect Rydberg population and perform feed-forward corrections. Several approaches can be considered. One promising way to rapidly measure individual qubit states is to resonantly  drive a cycling transition  and detect the scattered photons~\cite{Bergschneider18}. At lattice spacings of a few microns, this detection scheme could be limited by atom heating and  cross-talk from the reabsorption of scattered photons by neighboring atoms~\cite{Steck03}.  To this end, recent developments in coherent transport of entangled atom arrays~\cite{Bluvstein21b} can be used to mitigate these effects by moving the selected ancillary atom(s) into a detection zone far away from the rest of the array before it is measured. 

To estimate the maximum speeds of coherent transport before atom loss and heating become significant, one can consider the harmonic oscillator potential (i.e. the optical tweezer) that the atom is trapped in. 
Following the analysis in Refs.~\cite{Carruthers65, Bluvstein21b}, the average energy increase to the atom will be $\Delta E = m |\tilde{a}(\omega_0)|^2/2$, where $m$ is the particle mass and $\tilde{a}(\omega_0)$ is the Fourier transform of the acceleration profile $a(t)$ evaluated at the trap frequency $\omega_0$. When $a(t)$ is linear in time, this energy depends on the total displacement $D$ and the time of movement $T$ as approximately $\Delta E = 36 m  D^2 / (\omega_0^2 T^4).$
Based on this estimate, it is reasonable to achieve substantial atom displacements $D >50~\mu$m  within $250~\mu$s for performing feed-forward applications: for typical trap frequencies $\omega_0 \approx 2\pi \times 50 \textrm{~kHz}$, the atom's vibrational quantum number would increase by only $\Delta N < 1$. Indeed, such transport has been demonstrated by Ref.~\cite{Bluvstein21b} without significant decoherence or atom loss due to heating. Moving the atoms by a distance $D$ would then suppress reabsorption rates during ancilla readout to $\sigma/(4\pi D^2)$, where $\sigma$ is the absorption cross-section \cite{Steck03}. Moreover, detuning the optical transitions for ancilla atoms by $\Delta$ further suppresses reabsorption by a factor $\approx (\Gamma/2\Delta)^2$, where $\Gamma$ is the resonance linewidth, and $\Delta > 10 ~\Gamma$ can be readily achieved with moderate powers of a light-shifting beam \cite{Steck03}. Between moving and light-shifting the ancillary atoms, cross-talk errors on the data qubits can be suppressed by five or more orders of magnitude, to negligible levels.

Alternatively, the measurement of ancilla qubit states can be achieved by using two different atomic species for the data and ancilla atoms (such as two different isotopes of the same atom or two different atomic species). In this approach~\cite{Zeng17,Singh21}, the ancilla atoms can still interact with the data atoms when both are coupled to Rydberg states, while they can be measured independently without disturbing the data atom states.

Finally, fast detection schemes were recently demonstrated in experiments with small atomic ensembles using Rydberg electromagnetically induced transparency (EIT)~\cite{Xu21}. These  could be potentially utilized to identify Rydberg population after entangling gates. These schemes could be incorporated into the tweezer array platforms by creating larger, elongated traps at selected locations containing optically dense atomic ensembles. 
In this approach, the Rydberg blockade effect leads to a sharp signature in the absorption spectrum of a weak EIT probe beam depending on whether a nearby Rydberg atom is present. Due to the collectively enhanced Rabi frequency, the detection time can be reduced to $\sim 6 \mu\textrm{s}$~\cite{Xu21}, comparable to the duration of an entangling gate. This ultrafast, non-destructive Rydberg atom detector thus provides a promising implementation for the measurement and feed-forward corrections needed for our protocols.

\subsection{Implementation with alkaline earth(-like) atoms}

\begin{figure}
\includegraphics[width=0.8\textwidth]{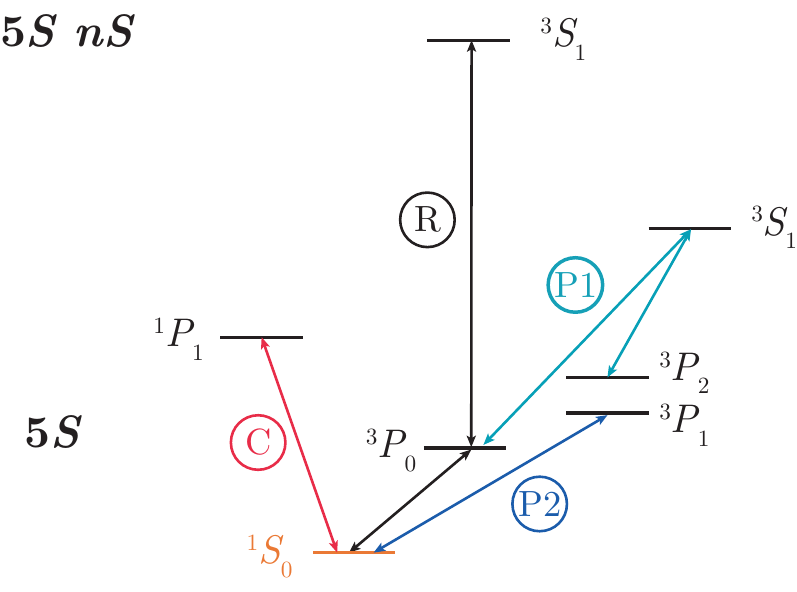}
\caption{Relevant level diagram for implementing our FTQC protocols with neutral alkaline earth Rydberg atoms such as $^{87}$Sr. The qubit is encoded in the stretched $^1S_0$ ground state (orange). Transitions to a $5S nS$, $^3S_1$ Rydberg state can be driven by first coherently mapping one of the qubit states to the $^3P_0$ clock state and then exciting the clock state to the Rydberg state 
(R, black). Optical pumping to correct for non-Rydberg leakage is implemented in two stages by driving the light blue transitions (P1) followed by the dark blue transition (P2). State readout and strong cooling for state initialization are implemented via the $^1S_0 \leftrightarrow {}^1P_1$ transition (C, red), while narrow-line cooling can be implemented via the P2 transition.}
\label{fig:alkaline-earth}
\end{figure}

In this work, we have focused primarily on developing FTQC protocols for neutral alkali atoms coupled to Rydberg states. Recently, significant progress has also been made towards using alkaline earth(-like) atoms such as Sr and Yb for Rydberg-based quantum computations \cite{Madjarov20,Wilson19}. In this section, we show how our methods can also be applied for such setups. While we focus on an example of $^{87}$Sr for concreteness, our discussion is generic for fermionic species of alkaline earth(-like) atoms.

For alkaline earth(-like) atoms, the $^1S_0$ ground states  have no electronic orbital or spin angular momentum, so the only source of degeneracy is the nonzero nuclear spin (which can be quite large, e.g.~$I = 9/2$ for $^{87}$Sr). 
For our protocols, a most convenient qubit encoding uses the stretched ground states: $|0 \rangle \equiv | m_I = -I  \rangle$, $|1 \rangle \equiv |m_I = +I \rangle$. In this encoding, strong cooling and state readout can be implemented via the $^1S_0 \leftrightarrow {}^1P_1$ transition, while narrow-line cooling can be performed on the $^1S_0 \leftrightarrow {}^3P_1$ transition. Entangling gates can be implemented by selectively exciting the $|1 \rangle$ state to a stretched Rydberg $^3S_1$ state. This state selectivity can be achieved by coherently mapping one of the qubit states to the $^3P_0$ clock state, performing Rydberg pulses between the clock state and the Rydberg state, and mapping back to the $^1S_0$ ground state, where we have utilized the linear Zeeman shift in the clock transition arising from hyperfine coupling between the $^3P_0$ and $^3P_1$ states~\cite{Boyd07}. 
The relevant level diagram is shown in Figure \ref{fig:alkaline-earth} for the case of $^{87}$Sr (see also Ref.~\cite{Ludlow15}).

During these entangling operations, an atom in the Rydberg state may undergo various errors such as BBR transitions, RD, or intermediate state scattering as described in Section \ref{sec:error-models}. For alkaline earth(-like) atoms, the resulting Kraus operators can be described by Pauli-$Z$ errors and quantum jumps to Rydberg states, $^1S_0$ ground states, or metastable $^3P$ states as allowed by dipole selection.

Following our approach for alkali atoms, we must convert all  such errors to Pauli-$Z$ errors to apply our FTQC protocols. By using ancilla atoms and the blockade effect, the quantum jumps to Rydberg states can be corrected in the same fashion as for alkali atoms. However, due to the presence of metastable $^3P$ levels, the correction of non-Rydberg leakage errors is more complicated, and the optical pumping must be done in two stages (see Figure \ref{fig:alkaline-earth}):
(1) Use $\sigma^+$-polarized light from the $^3P_{0,2}$ states to the triplet excited $^3S_1$ state to re-pump all $^3P$ states to the $^3P_1$ manifold; these states will decay back into the $^1S_0$ ground states.
(2) Use $\sigma^+$-polarized on the narrow-line cooling transition $^1S_0 \leftrightarrow {}^3P_1$ to pump ground states with $m_F > -I$ to the stretched ground state $|1 \rangle = |m_I = +I \rangle$. 
After these two steps, all non-Rydberg leakage errors will be mapped to the error $|1 \rangle \langle 1|$, which is expressible in terms of Pauli-$Z$ errors. We note that while Pauli-$X$ errors could in principle arise from polarization impurities in the $^1S_0 \leftrightarrow {}^3P_1$ beam in the second stage, this would require several consecutive polarization imperfections, each of which has a very low probability of roughly $0.2-0.5\%$; thus, the overall probability of Pauli-$X$ errors arising from imperfect polarization is negligible.
Therefore, by using this optical pumping scheme to convert all non-Rydberg leakage errors to $Z$ errors, the FTQC schemes of Sections \ref{sec:ftqc-7} and \ref{sec:ftqc-repetition} can be implemented in alkaline earth(-like) atoms.

\section{Conclusions and Outlook}
\label{sec:conclusions}

In this work, we have presented a detailed analysis of the dominant error channels arising in quantum computation using neutral Rydberg atoms. We show that although the multilevel nature of atoms and the complex decay channels for Rydberg states lead to many additional types of errors not considered in traditional QEC settings, the specific structure of the error model allows us to design hardware-efficient FTQC protocols based on the seven-qubit and hardware-tailored three-qubit codes with significantly reduced overhead compared to general-purpose schemes.  The crux of these results is the ability to convert the complicated error model to Pauli-$Z$ errors by introducing ancilla atoms and making use of the Rydberg blockade effect, dipole selection rules, and new schemes for optical pumping. To use the three-atom repetition code, we designed a new laser pulse sequence to implement bias-preserving CNOT and Toffoli gates. For both protocols, we propose simple, scalable geometrical layouts and demonstrate feasibility of all components of FTQC for near-term implementation.

Our results provide an important step towards building large-scale quantum computers based on neutral atom setups and highlight the importance of designing FTQC schemes based on the specific structure of the error model and the unique capabilities of the hardware platform. Compared to some general-purpose FTQC protocols, our hardware-efficient approaches for Rydberg systems enable an order-of-magnitude improvement in resource overhead in terms of the number of physical gates or required ancillas. We believe many of the ideas developed in this work, such as the exploitation of the multi-level structure of the physical system, are transferable to other quantum computing platforms such as trapped ions and superconducting qubits. In the former case, an optical pumping-based protocol to mitigate leakage in ions with low nuclear spin such as $^{171}$Yb$^+$ was recently developed and realized experimentally~\cite{Hayes20}; we believe that insights from our work would be helpful for developing more general leakage correction methods in such setups and incorporating them into FTQC protocols.

Several interesting extensions can be considered. For example, while we have primarily quantified the resource cost for FTQC proposals by studying the number of qubits and gates required, another related and commonly used metric is the error threshold, which amounts to the physical qubit and gate fidelities required to produce logical error rates that are lower than the physical error rate. One may estimate the error thresholds for two- and three-qubit gates directly by using the numbers presented in Tables~\ref{tab:stab-comparison} and \ref{tab:gate-comparison} and requiring each logical operation or stabilizer measurement step to have at most a single error, but it can be more useful to obtain the precise numbers for these thresholds via numerical simulation.

A more detailed study of the error threshold would be especially helpful if one intends to extend our work to codes with distance greater than $3$ and compare the relative performance and scalability of these approaches with our current proposal. One particularly intriguing direction could be to evaluate the performance and resource cost of existing FTQC protocols based on topological codes such as surface codes or color codes~\cite{Bravyi98,Fowler12,Bombin06,Bombin07,Terhal15,Auger17} upon applying our techniques to address Rydberg and non-Rydberg leakage errors. Indeed, one recent work has already demonstrated ultrahigh error threshold in the surface code when the underlying noise model is biased~\cite{Tuckett18}; this motivates the use of surface codes in a Rydberg system where dominant errors have all been converted to Pauli-$Z$ type. For near-term implementations, more detailed studies considering a combination of current experimental capabilities and specific technical imperfections together with the intrinsic Rydberg state decay errors may also allow for improved error rates in encoded qubits and operations. Finally, after eliminating all of the Rydberg-specific leakage errors using our FTQC protocols, one could concatenate our codes with more traditional QEC approaches to address any higher-order Pauli-$X$ or $Y$-type errors that were neglected in our studies, or to further suppress the logical error rate to even higher orders.

\vspace{3mm}
\emph{Note added}: 
In the process of revising and preparing this second version of our manuscript, we became aware of the related work of Ref.~\cite{Wu22}, which proposes an efficient approach to FTQC in alkaline earth atoms based on converting Rydberg leakage into erasure errors by detecting the exact location of the leakage and replacing atoms lost during the detection process.

\vspace{3mm}
\emph{Acknowledgments}: We thank Maya Miklos, Soonwon Choi, Florentin Reiter, Hannes Pichler, and Hyeongrak Choi for helpful discussions. Special thanks go to Ofer Firstenberg and Leo Zhou for providing code which we used to calculate dipole matrix elements for the blackbody radiation and radiative decay transitions in $^{87}$Rb. 
This work was supported by the Center for Ultracold Atoms, the National Science Foundation, the U.S. Department of Energy (DE-SC0021013 \& LBNL QSA Center), ARO MURI, the DARPA ONISQ program, QuEra Computing, and Amazon Web Services . I.C. acknowledges support from the Alfred Spector and Rhonda Kost Fellowship of the Hertz Foundation and the Paul and Daisy Soros Fellowship. I.C. and H.L. acknowledge support from the Department of Defense through the National Defense Science and Engineering Graduate Fellowship Program. D.B. acknowledges support from the NSF Graduate Research Fellowship Program (grant DGE1745303) and the Susan and Richard Miles Fellowship of the Hertz Foundation.

\begin{appendix}

\section{Numerical Computation of Branching Ratios and Transition Rates}
\label{sec:transition-rates-app}

In this section, we present the results of numerical computation of branching ratios for BBR and RD transitions out of the stretched Rydberg state $70S_{1/2}$, $m_J=1/2$, $m_I=3/2$ for $^{87}$Rb.

\subsection{Blackbody radiation-induced transitions}

\begin{figure}
\includegraphics[width=0.8\textwidth]{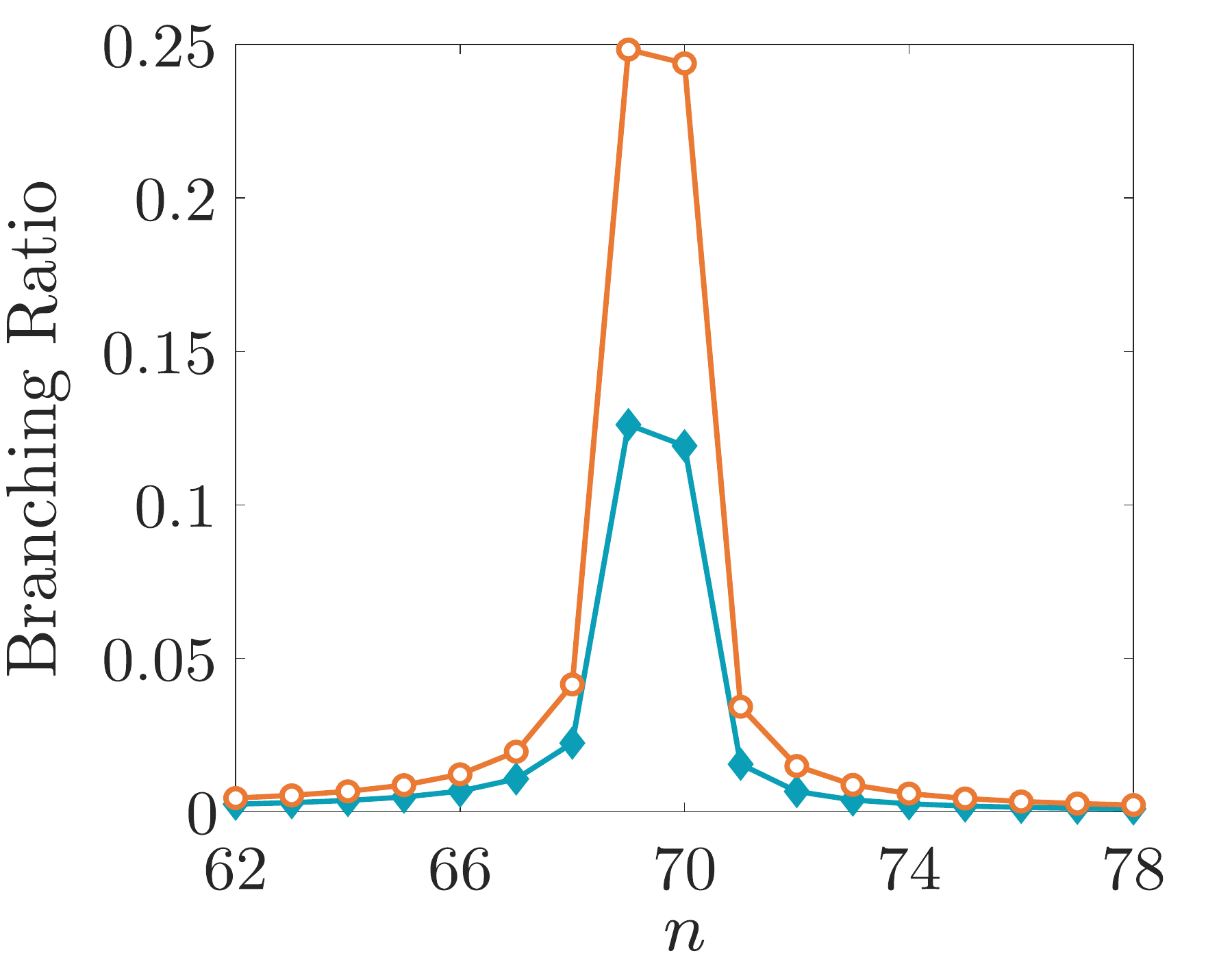}
\caption{Branching ratios for BBR transitions between Rydberg states of $^{87}$Rb, from the stretched $70S_{1/2}$ state with $m_J=1/2$, $m_I=3/2$ to different $P$ states with $m_J=3/2$  (empty orange circles) or $m_J=1/2$ (filled blue diamonds).}
\label{fig:bbr-70s}
\end{figure}

To quantify the relative probability of transitioning into different nearby Rydberg $P$ states, we compute the rate $W(nL \rightarrow n'L')$ of BBR transitions from a given Rydberg state $nL$ to other Rydberg states $n'L'$ using the Planck distribution of photons at the given temperature $T$ and the Einstein coefficient for the corresponding transition:
\begin{equation}
\label{eq:bbr-rate}
W(nL \rightarrow n'L') = A(nL \rightarrow n'L') \bar{n}_\omega
\end{equation}
\noindent
where $\omega = E_{nL} - E_{n'L'}$ is the transition frequency ($E_{nL}$ and $E_{n'L'}$ are energies of the initial and final states) and
\begin{equation}
\label{eq:einstein-A}
A(nL \rightarrow n'L') = \frac{4\omega^3}{3c^3} \frac{L_{\textrm{max}}}{2L+1} R^2(nL \rightarrow n'L').
\end{equation}
In the above equations, $\hbar = 1$, $L_\textrm{max} = \textrm{max}(L,L')$, and $R(nL \rightarrow n'L')$ is the radial matrix element for the electric dipole transition $nL \rightarrow n'L'$. 

For this work, we used analytic formulas from Refs. \cite{Kaulakys95} and \cite{Kamta98} to numerically compute the radial dipole matrix elements for single-photon BBR transitions from the stretched Rydberg state $70S_{1/2}$, $m_J=1/2$, $m_I=3/2$ of $^{87}$Rb. We then computed the corresponding transition rates using Eq.~(\ref{eq:bbr-rate}), and normalized these by the total BBR rate
$\Gamma_{\textrm{BBR}}$ (see Eq.~(\ref{eq:bbr-scaling})) to obtain the branching ratios.  
The branching ratios for 
$P$ states with $m_J=3/2$ and $m_J=1/2$ are plotted in Figure \ref{fig:bbr-70s} as empty orange circles and filled blue diamonds, respectively. Indeed, we find that the atom decays primarily to the $69P$ and $70P$ states as illustrated in Figure~\ref{fig:overview-figure}c.

\subsection{Radiative decay}

As shown in Figure \ref{fig:overview-figure}c of the main text, the radiative decay transitions from the stretched $70S_{1/2}$, $m_J=1/2$, $m_I=3/2$ Rydberg state of $^{87}$Rb are almost entirely two- or four-photon decay processes to one of the five states in the ground state manifold; this fact was important for converting all Rydberg errors to $Z$ type for fault-tolerant quantum computation. To justify this, we numerically computed the branching ratios for multi-photon spontaneous emission processes  by evaluating the ratios of individual transition rates for each decay channel, which are given by the Einstein $A$ coefficients of Eq.~(\ref{eq:einstein-A}). Due to the cubic dependence of these coefficients on transition frequency, the primary contributions arise from dipole-allowed transitions to states near the ground state manifold. The dipole matrix elements for such transitions scale with the effective principal quantum number $n_{\textrm{eff}}$ of the Rydberg state as $\sim 1/n_{\textrm{eff}}^{1.5}$. The total RD rate is then given by a sum over Einstein coefficients for all possible target states:
\begin{equation}
\label{eq:radiative-decay-rate}
\frac{1}{\tau_0} = \Gamma_0 = \sum_{n'L': E_{nL} > E_{n'L'}} A(nL \rightarrow n'L').
\end{equation} 
By computing the radial dipole matrix elements using analytic formulas from Refs.~\cite{Kaulakys95} and \cite{Kamta98}, we evaluated the branching ratios for RD processes out of the $70S_{1/2}$, $m_J=1/2$, $m_I=3/2$ stretched Rydberg state for $^{87}$Rb.

\begin{table}[tbp!]
\caption{Branching ratios for transition to each ground state of $^{87}$Rb for radiative decay processes from the $70S_{1/2}$, $m_J=1/2$, $m_I=3/2$ stretched Rydberg state, accounting for transitions involving up to four-photon emission processes. The contribution from transitions of even higher order is less than $2.5 \times 10^{-4}$.}
\label{tab:70s-rd-branching}
\begin{tabular}{|c|c|c|} \hline
 $\bm{F}$ &  $\bm{m_F}$ & {\bf Branching ratio} \\ \hline
2 & 2 & 0.534 \\ \hline
2 & 1 & 0.177 \\ \hline
2 & 0 & 0.055 \\ \hline
2 & $-1$ & 0.003 \\ \hline
2 & $-2$ & 0.001 \\ \hline
1 & 1 & 0.168 \\ \hline
1 & 0 & 0.059 \\ \hline
1 & $-1$ & 0.003 \\ \hline
\end{tabular}
\end{table}

The results of this computation are shown in Table \ref{tab:70s-rd-branching}. Indeed, we find that the branching ratios for the remaining three states are each on the order of $10^{-3}$, significantly smaller than those for the dominant five transitions. If the total error probability is already very small, these three processes (in particular, the decay to the stretched state with minimal $m_F=-2$) are highly unlikely. 

\section{An Example of Master Equation Solution for Radiative Decay}
\label{sec:master-eq-app}

In Section \ref{sec:radiative-decay}, we argued that the Kraus operators corresponding to spontaneous emission events from the Rydberg state $|r\rangle$ to the qubit $|1 \rangle$ are 
\begin{equation}
\label{eq:rd-kraus}
\begin{gathered}
M_0 = |r \rangle \langle r| + \alpha |1 \rangle \langle 1| + \beta |0 \rangle \langle 0|, \\
M_{r} \propto |r \rangle \langle 1|, \quad
M_{1} \propto |1 \rangle \langle 1|, \quad 
M_{2} \propto  |0 \rangle \langle 0|,
\end{gathered}
\end{equation}
where $\alpha$, $\beta$, and the proportionality constants depend on the specific Rydberg pulse being performed and the probability for an atom in the Rydberg state to decay to the $|1 \rangle$ state. We now proceed to derive these constants for the special case of a $2\pi$ pulse on the Rydberg transition $|1 \rangle \leftrightarrow |r \rangle$ by analytically solving the quantum master equation. For this example calculation, we will ignore BBR transitions and RD transitions to other hyperfine states; these can be included as a straightforward extension.

The master equation for this driven three-level system is (setting $\hbar = 1$)
\begin{equation}
\label{eq:master}
\frac{d\hat{\rho}}{dt} = -i[\hat{H_d},\hat{\rho}] - \frac{\gamma}{2}\left(\hat{c}^\dagger \hat{c} \hat{\rho} + \hat{\rho} \hat{c}^\dagger \hat{c} - 2 \hat{c} \hat{\rho} \hat{c}^\dagger \right),
\end{equation}
where $\hat{\rho}$ denotes the density matrix of the system, $\hat{H_d} = i \Omega \left(|r\rangle \langle 1| - |1 \rangle \langle r| \right)$ is the driving Hamiltonian, $\hat{c} = |1 \rangle \langle r |$ is the quantum jump operator corresponding to spontaneous emission $|r \rangle \mapsto |1 \rangle$, and $\gamma$ is the probability for an atom in the Rydberg state to decay to $|1 \rangle$. We assume the qubit is initially encoded in the hyperfine manifold $\textrm{Span}\{|0 \rangle, |1 \rangle \}$, so that the initial density matrix can be written as
\begin{equation}
\hat{\rho}_0 = 
\begin{bmatrix}
0 & 0 & 0 \\
0 & \rho_{11} & \rho_{10} \\
0 & \rho_{01} & \rho_{00}
\end{bmatrix}
\end{equation}
(we order the matrix columns and rows as $\{|r \rangle, |1 \rangle, |0 \rangle\}$). Upon solving the resulting coupled first-order differential equations, we find that the final state after the $2\pi$ pulse with decay is, to leading order in $\gamma/\Omega$,
\begin{equation}
\label{eq:rho-final-2pi}
\hat{\rho}_f = 
\begin{bmatrix}
3 \gamma t_\pi \rho_{11}/4 & 0 & 0 \\
0 & (1-3 \gamma t_\pi/4) \rho_{11} & -e^{-\gamma t_\pi /2} \rho_{10} \\
0 & -e^{-\gamma t_\pi /2} \rho_{01} & \rho_{00} 
\end{bmatrix}.
\end{equation}
Here, $t_\pi = \frac{\pi}{2\Omega}$ is the duration of a $\pi$ pulse. Indeed, Eq. (\ref{eq:rho-final-2pi}) confirms our intuition from Section \ref{sec:radiative-decay} that the coherences $\rho_{r1}$, $\rho_{1r}$ vanish upon ``averaging'' over all possible transition times during the $2\pi$ pulse.

One can then verify that, to leading order in $\gamma/\Omega$, the Kraus operators
\begin{equation}
\begin{aligned}
M_0 = & |r \rangle \langle r| + \sqrt{1-p_2} |0 \rangle \langle 0|\\
& + \sqrt{(1-p_1)(1-p_2)} |1 \rangle \langle 1| 
\end{aligned}
\end{equation}
\begin{equation}
M_r = \sqrt{p_1(1-p_2)} |r \rangle \langle 1|
\end{equation}
\begin{equation}
M_1 = \sqrt{p_2} |1 \rangle \langle 1|
\end{equation}
\begin{equation}
M_2 = \sqrt{p_2} |0 \rangle \langle 0|
\end{equation}
give rise to the desired evolution from $\rho_0$ to $\rho_f$ provided we take $p_1 = 3 \gamma t_\pi/4$ and $p_2 = \gamma t_\pi / 8$.

\section{Atom Loss Errors}
\label{sec:atom-loss-app}

\begin{figure}[!t]
\includegraphics[width=0.5\textwidth]{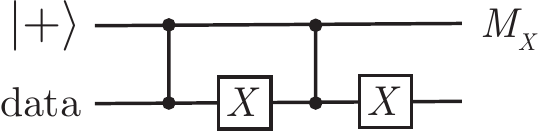}
\caption{Circuit for detecting atom loss.}
\label{fig:leakage-detection}
\end{figure}

As mentioned in Section \ref{sec:other-errors} of the main text, neutral atom setups can also suffer from atom loss errors if the trapping is imperfect, or if the trapping lasers need to be turned off during Rydberg excitation (e.g.\ as done for $^{87}$Rb in Ref.~\cite{Bernien17}). Fortunately, such errors can also be detected and corrected within our FTQC framework at the cost of one ancilla qubit and some extra gates for each operation. In particular, an atom loss event can be detected by applying the circuit of Figure~\ref{fig:leakage-detection} for each data qubit after using the optical pumping technique to correct for leakage out of the computational subspace. The ancilla measurement will then produce $+1$ in the presence of atom loss, and $-1$ if such an error did not occur. Once detected, an atom loss error can be converted to a single-qubit Pauli-$Z$ or $X$ type error if a reservoir of atoms is available, for instance by replacing the lost atom with a new atom initialized to the $\ket{0}$ state. 

We now discuss the steps needed for establishing robustness against errors occurring during this circuit. As in the case of fault-tolerant Rydberg leakage detection discussed in Appendix~\ref{sec:rydberg-leakage-app}, to protect against ancilla errors in Figure \ref{fig:leakage-detection}, we again adopt a multi-step ancilla measurement protocol  and require two positive ancilla measurements to confirm an atom loss error. On the other hand, any phase-flip error on the data qubit cannot propagate to more than a single physical qubit error per logical qubit in the universal gate set implementations for Ryd-7 or Ryd-3. Leakage errors (Rydberg or non-Rydberg) can be addressed by repeating the respective re-pumping procedures after applying the atom loss detection circuit. 
Thus, by incorporating this circuit into the implementation of fault-tolerant stabilizer measurements and logical operations in Sections \ref{sec:ftqc-7} and \ref{sec:ftqc-repetition}, we can also address atom loss errors in our FTQC protocols.

Notice that this circuit can be used for atom loss after correcting for leakage into atomic states outside the computational subspace by using the blockade effect and optical pumping techniques. In addition, this approach does not distinguish between atom loss and leakage into other hyperfine states, so it can also be used to suppress any residual hyperfine leakage errors.

\section{Converting Rydberg Leakage to Pauli Errors}
\label{sec:rydberg-leakage-app}

Once a Rydberg leakage error is detected, it can be converted to an atom loss error by ejecting the Rydberg atom, which is naturally done by the anti-trapping potential from the tweezer~\cite{Bluvstein21b} and can be expedited by pulsing a weak, ionizing electric field~\cite{Beguin13,Cohen03,Wu22}. The exact location of the ejected atom can be determined by following the atom loss protocol outlined in Appendix~\ref{sec:atom-loss-app} and Figure~\ref{fig:leakage-detection}~\footnote{In this case, the atom loss protocol does not need to be applied in a robust fashion, since an error has already occurred.}; subsequently, the ejected atom can be replaced with a fresh atom prepared in the $|1\rangle$ state.
Although this process simply replaces the Rydberg atom by an atom in the $\ket{1}$ state, by using the operator identity $|1 \rangle \langle 1| = \frac{1}{2}\left(\one - Z\right)$, it is straightforward to see that the resulting state is now a superposition of the original state without error, and the same state with a $Z$ error on this physical qubit. Such $Z$-type errors can be detected and corrected for using stabilizer measurements in both the seven-qubit and three-qubit codes. This procedure can also be modified to convert the Rydberg leakage error to a Pauli $X$-type error by applying Hadamard gates at the beginning and end; this is used to in the logical $\CCZ$ gate for Ryd-7 (see Algorithm~\ref{alg:7q-ccz}).

To reduce the need for applying the atom loss correction circuit, one could add a preventative step after every entangling gate which incoherently re-pumps any remnant population in several most probable Rydberg states into the qubit $|1 \rangle$ state. This re-pumping can be implemented via the following three-step procedure: 
\begin{enumerate}
\item Swap the population in $|1 \rangle$ and the stretched ground state $|F=I+1/2,m_F=I+1/2 \rangle$. 
\item For the most probable final states $\ket{r'}$ of a BBR transition (or the Rydberg state $\ket{r}$ in the case of RD), perform a Rydberg laser pulse that sends $\ket{r'}$ (or $\ket{r}$) to a short-lived $P$ state. In particular, we choose the $P$ state with the smallest possible $n$, largest possible $F$, and largest possible $m_F$. This state will quickly decay to the stretched state $|F=I+1/2,m_F=I+1/2 \rangle$, and cannot decay to any other ground state.
\item Repeat Step (1).
\end{enumerate}
By applying this procedure preventatively, one can convert a large fraction of Rydberg leakage errors to $Z$-type errors without the need for the atom loss correction circuit of Figure~\ref{fig:leakage-detection}.

\section{Fault-Tolerant Detection of Rydberg Leakage Errors}
\label{sec:ancilla-multi-step-app}

As mentioned in the main text, for fault-tolerant error detection and correction, it is important to address any errors that may occur on an ancilla used to probe for Rydberg population. This can be done by using a multi-step measurement procedure to detect leakage for the ancilla qubit:
\begin{enumerate}
\item Perform a Hadamard gate on the ancilla.
\item Check whether the ancilla is in the $\ket{1}$ state (e.g.\ by coupling $\ket{1}$ to a cycling transition and detecting fluorescence).
\item Perform an $X$ gate on the ancilla.
\item Check for $\ket{1}$ population again.
\end{enumerate}
If neither the second nor the last step yields $\ket{1}$, the ancilla atom must have undergone a leakage error. In that case, we convert any possible ancilla atom Rydberg error to a possible $Z$-type error as described in Appendix \ref{sec:rydberg-leakage-app}. Similarly, because the Rydberg pulses can potentially cause a phase-flip error on the ancilla qubit, if a Rydberg leakage error is detected by the ancilla, the detection protocol must be repeated once more to ensure that the outcome did not result from such an error.

\section{Error Syndromes with Postponed 
Measurements}
\label{sec:postponing-app}

In Section \ref{sec:ryd-bbr}, we discussed how Rydberg leakage detection can be postponed in the Ryd-7 stabilizer measurement and controlled-phase gate protocols to facilitate experimental implementation. This relied on the ability to use stabilizer measurements to distinguish between the possible correlated errors that can result from postponed detection of a Rydberg leakage error. Here, we present details on how to use error syndromes to identify the corresponding correlated error in each case. As in the main text, we assume the stabilizers for the Steane code are ordered as
\begin{align}
\label{eq:steane-stabilizer-app}
&g_1 = IIIXXXX  & g_2& = IXXIIXX  & g_3 &= XIXIXIX \nonumber \\
&g_4 = IIIZZZZ & g_5& = IZZIIZZ  &g_6& = ZIZIZIZ.
\end{align}

\begin{table}[t]\caption{Using error syndromes to distinguish between correlated errors resulting from postponed detection of Rydberg leakage during measurement of the $X_4 X_5 X_6 X_7$ stabilizer in the Ryd-7 FTQC protocol. Because the possible correlated errors are all products of Pauli-$X$ errors, we show the corresponding values of $Z^{\otimes 4}$ stabilizer measurements.}
\label{tab:stabilizer-postponed}
\begin{tabular}{|c|c|c|c|}
\hline {\sc Error} & $g_4$ & $g_5$ & $g_6$ \\ \hline
$X_5 X_6 X_7$ & $-1$ & $+1$ & $+1$ \\ \hline
$X_6 X_7$ & $+1$ & $+1$ & $-1$ \\ \hline
$X_7$ & $-1$ & $-1$ & $-1$ \\ \hline
\end{tabular}
\end{table}

For the stabilizer measurement, we will consider (without loss of generality) the measurement of $g_1$ on qubits $4,5,6,7$ using a circuit of the form shown in Figure~\ref{fig:overview-figure}b. If a Rydberg leakage error occurs on the ancilla atom at any point, the data atoms do not suffer any correlated errors. On the other hand, if a data atom suffers a Rydberg leakage error during the circuit, the possible correlated errors that can result are $X_5 X_6 X_7$, $X_6 X_7$, or $X_7$. These errors can be distinguished by measuring the $Z^{\otimes 4}$ stabilizers of the seven-qubit code; the corresponding error syndromes are shown in Table \ref{tab:stabilizer-postponed}.

For the case of the logical $\CCZ$ gate, we grouped the 27 physical Rydberg gates into groups $\mathcal{G}_i$ of three, and performed Rydberg leakage detection after each group. Without loss of generality, we will consider the group $\mathcal{G}_1$ in Figure~\ref{fig:ccz-reordering}. There are two possible correlated errors that could result from the delayed detection of Rydberg leakage in this case (up to a single-qubit error within each logical qubit): $R(2_A,2_B;2_C) R(3_A,3_B;3_C)$ and $R(3_A,3_B;3_C)$. By writing the Rydberg gate as
\begin{equation}
    R(j,k;l) = (\one + Z_j) (\one + Z_k) (\one + Z_l) - \one,
\end{equation}
we find that the two cases can be distinguished by measuring the stabilizers 
$g_2$ and $g_3$ for each of the logical qubits. In the former case, at least one of the the logical qubits would have either a $Z_2$ or $Z_2 Z_3$ error, giving rise to stabilizer eigenvalues $(g_2, g_3) = (-1, +1)$ or $(+1,-1)$, while in the latter scenario, all three sets of stabilizer measurements would yield $(-1,-1)$ or $(+1,+1)$.

\section{Implementation of a Bias-Preserving Toffoli Gate}
\label{sec:toffoli-app}

\begin{figure*}[!t]
\includegraphics[width=0.75\textwidth]{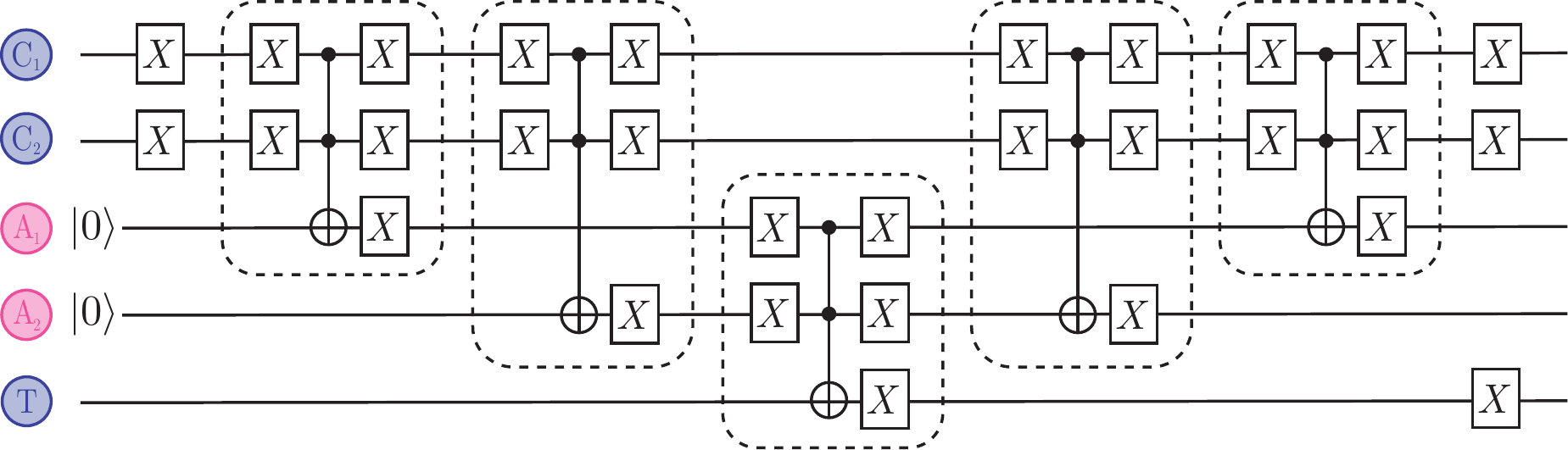}
\caption{Using two ancilla qubits and multiple Rydberg states to implement a bias-preserving Toffoli gate between control atoms $C_1$, $C_2$, and target atom $T$. The ancilla atoms ($A_1$ and $A_2$) are chosen to lie on either side of the target atom. The dotted boxes indicate the most natural bias-preserving three-qubit gate for Rydberg systems, where $\pi$ pulses $|1\rangle \leftrightarrow |r_+\rangle$ are applied to each of the first two (i.e., the upper two) involved atoms, the bias-preserving pulse sequence of Figure~\ref{fig:bias-preserving-cnot-target} is applied to the third (lower) atom, and $-\pi$ pulses $|1\rangle \leftrightarrow |r_+\rangle$ are applied to the first two qubits; the Rydberg states $|r_\pm \rangle$ are chosen to be either $|r_{1,\pm}\rangle$ or $|r_{2,\pm} \rangle$ for each such gate. In this circuit, we set $|r_\pm \rangle = |r_{1,\pm}\rangle$ in the first, second, fourth, and fifth cases, while choosing $|r_\pm \rangle = |r_{2,\pm}\rangle$ for the third one. With this choice of Rydberg levels, the two ancillas will not interact with each other during the third Rydberg gate. However, we note that the two control atoms may interact with each other during the first, second, fourth, and fifth entangling gates if the distance between them is less than one blockade radius; this is not problematic because Rydberg errors can occur during at most one of these gates, so at least one ancilla atom will generate the correct interaction with the target atom during the third gate.}
\label{fig:bias-preserving-toffoli}
\end{figure*}

In Figure~\ref{fig:bias-preserving-cnot-control}, we showed how an ancilla atom can be used to eliminate $X$-type errors resulting from control atom decay in the implementation of a bias-preserving CNOT gate. Analogously, a bias-preserving Toffoli gate can be implemented by making use of two ancilla atoms which lie on either side of the target atom. This protocol is  illustrated in Figure~\ref{fig:bias-preserving-toffoli}.

As with the case of the bias-preserving CNOT, the choice of Rydberg states differs throughout the procedure. By coupling the atoms to $|r_{2,\pm}\rangle$ during the third gate of Figure~\ref{fig:bias-preserving-toffoli} and using ancilla atoms on opposite sides of the target atom, we ensure that the ancilla atoms do not interact with each other via Rydberg blockade during this gate; this is important in case one of the ancilla atoms undergoes a radiative decay transition during this gate. On the other hand, the other entangling gates in Figure~\ref{fig:bias-preserving-toffoli} all use the Rydberg states $|r_{1,\pm}\rangle$, due to larger distances between the atoms during these gates. We note that the two control atoms may interact with each other during these four gates if the distance between them is less than one blockade radius, which is different from the case of the third gate. This is acceptable because Rydberg errors can occur during at most one of these four gates, so at least one ancilla atom will generate the correct interaction with the target atom during the third gate.

\section{Computing Resource Costs for Rydberg FTQC Protocols}
\label{sec:resource-cost-app}

We now provide details on how to obtain the resource costs for Ryd-7 and Ryd-3 presented in Tables \ref{tab:stab-comparison} and \ref{tab:gate-comparison}.

For the Ryd-7 protocol, each stabilizer measurement requires four two-qubit Rydberg gates in the absence of errors (see Algorithm~\ref{alg:7q-x-stab}); thus, 24 two-qubit gates are required to measure all stabilizers. If an error occurs, the worst case scenario for the stabilizer measurement is when the first five stabilizers all have $+1$ eigenvalues, while the very last stabilizer is measured to be $-1$. In this case, $g_4$, $g_5$, and $g_6$ need to be re-measured,  which requires 12 additional two-qubit gates. The logical CCZ gate for Ryd-7 is implemented using 27 physical three-qubit gates in the absence of error, as described in Algorithm~\ref{alg:7q-ccz}. The worst case error in this case is a Rydberg leakage error that occurred during the first entangling gate in the final group $\mathcal{G}_9$ of Figure~\ref{fig:ccz-reordering}. In this scenario, identifying the location of the Rydberg leakage error requires up to 18 additional two-qubit gates, while measuring the stabilizers $g_2, g_3, ..., g_6$ for all three logical qubits would amount to 60 additional two-qubit gates; the correction circuit could require up to two additional three-qubit gates.

In the Ryd-3 protocol, each of the two stabilizer measurements requires two bias-preserving CNOT gates (Figure~\ref{fig:overview-figure}e), and each bias-preserving CNOT gate is broken down to two two-atom gates and one three-atom entangling gate (see Section~\ref{sec:rydberg-bias-preserving}). Thus, in the absence of error, the stabilizer measurements would require eight two-qubit gates and four three-qubit gates. If an error occurs, the worst case scenario is if the second stabilizer is measured to be $-1$; in this case, both stabilizers need to be re-measured, and the gate cost is doubled. The Ryd-3 $\CCZ$ gate can be implemented in a round-robin fashion in the same way as the Ryd-7 $\CCZ$, which is bias-preserving and uses 27 physical three-qubit gates.

\begin{figure}[!t]
\includegraphics[width=0.75\textwidth]{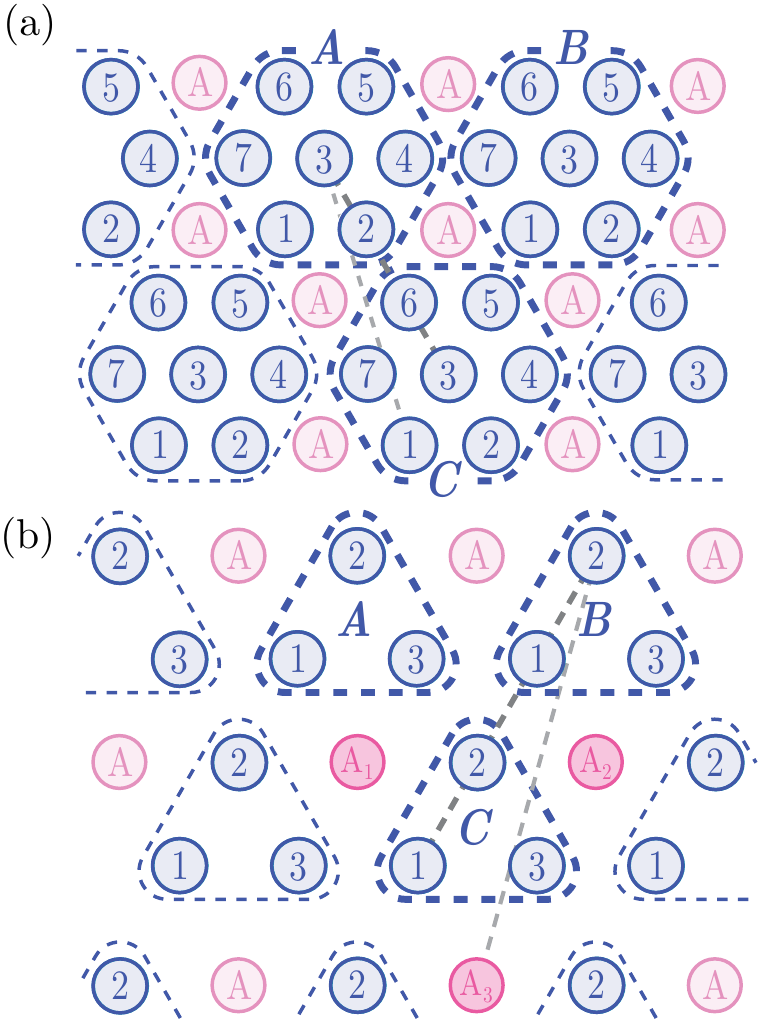}
\caption{Example labeling of atoms for the Ryd-7 and Ryd-3 FTQC protocols used to derive the gate counts and blockade radius requirements. As in Figure~\ref{fig:overview-figure}, data atoms are shown in blue, while ancilla atoms are shown in pink. (a) In the Ryd-7 protocol, each logical qubit consists of seven data atoms (blue dotted hexagons). For each data atom, a number is used to indicate which physical qubit of the seven-qubit logical state the atom encodes. With this labeling, the blockade radius $R_B$ is defined by the interaction range needed to perform a logical $\CCZ$ gate between three neighboring logical qubits such as $A$, $B$, and $C$. Using the specific $\CCZ$ protocol given in Algorithm~\ref{alg:7q-ccz}, the blockade radius requirement is then $R_B > 3.61d$, where $d$ is the spacing between nearest neighbors on the lattice; this is determined by the distance between physical atoms $3_A$ and $1_C$ (thinner, light grey dotted line). However, by using a different set of physical $\CCZ$ gates to implement the logical $\CCZ$, this requirement can be reduced to $R_B > 3d$ (thicker, dark grey dotted line). (b) In the Ryd-3 protocol, each logical qubit consists of three data atoms (blue dotted triangles). For each data atom, a number is used to indicate which physical qubit of the three-qubit logical state the atom encodes. With this labeling, the larger blockade radius $R_{B,1}$ is determined by the interaction range required for performing a logical Toffoli gate between three neighboring logical qubits such as $A$, $B$, and $C$. In this case, there are two possibilities for $R_{B,1}$---either $R_{B,1} > 3.61d$ (thinner, light grey dotted line) or $R_{B,1} > 3d$ (thicker, dark grey dotted line).  When the larger blockade radius of $3.61d$ can be realized, the resource cost for the logical Toffoli and Hadamard gates can be reduced by four two-qubit entangling gates compared to the numbers presented in Table~\ref{tab:gate-comparison} (see also Appendix~\ref{sec:blockade-app}).}
\label{fig:logical-qubit-numbered}
\end{figure}

Finally, the Ryd-3 Hadamard gate consists of a fault-tolerant, bias-preserving Toffoli gate followed by single-qubit measurements and rotations (Figure~\ref{fig:repetition-hadamard}). The pieceable fault-tolerant Toffoli gate in the Ryd-3 code consists of nine physical bias-preserving Toffoli gates and two rounds of error correction. As discussed above, each round of error correction involves eight two-atom Rydberg gates and four three-atom Rydberg gates. When the data atoms within each logical qubit are indexed as in Figure~\ref{fig:logical-qubit-numbered}b and we are implementing a logical Toffoli gate $\textrm{CCX}_{ABC}$ between the three qubits $A$, $B$, $C$ highlighted in bold, the number of Rydberg gates required to implement each physical Toffoli gate depends on the blockade radius $R_{B,1}$. If the blockade radius $R_{B,1}$ is larger than $3.61d$, each physical Toffoli gate can be implemented using two ancilla atoms (one on either side of the target atom) and five three-atom Rydberg gates, as described in Appendix~\ref{sec:toffoli-app}; this is because the distance between any physical control atom $C_i$ and any ancilla $A_j$ in Figure~\ref{fig:bias-preserving-toffoli} will always be less than the blockade radius $R_{B,1}$, so the entangling gates can be implemented directly. In this case, each physical Toffoli gate involves five three-atom Rydberg gates, so the total gate count (upon including the QEC steps) is 16 two-atom gates and 53 three-atom gates in the absence of errors. On the other hand, if we wish to reduce the blockade radius requirement to $R_{B,1} > 3d$, there are two physical Toffoli gates (corresponding to the choices $j_A = l_C = 1$, $k_B = 2$ and $j_A = l_C = 3$, $k_B = 2$), where the distance between one of the physical control atoms and one of the ancilla atoms ($2_B$ and $A_3$ in Figure~\ref{fig:logical-qubit-numbered}b) would be too large to directly implement a Rydberg entangling gate required for the physical Toffoli gate. Instead, in place of the first (respectively, second) three-atom Rydberg gate involving $A_3$, we would implement a Rydberg gate with the same two control atoms and one of the ancilla atoms $A_1$ or $A_2$, whichever is not involved in the rest of the Figure~\ref{fig:bias-preserving-toffoli} circuit, followed (respectively, preceded) by a bias-preserving CNOT gate between that ancilla and $A_3$. These gates can be implemented directly because both $A_1$ and $A_2$ are within the blockade radius of $2_B$, $1_A$, $2_A$, $3_A$, and $A_3$. In this way, four extra two-atom gates are required for the logical Toffoli (two for the physical Toffoli with $j_A = l_C = 1$, $k_B = 2$ and two for the physical Toffoli with $j_A = l_C = 3$, $k_B = 2$), which increases the total gate count to 20 two-atom gates and 53 three-atom gates in the absence of error, as shown in Table~\ref{tab:gate-comparison}. With errors, the worst case scenario is if the final stabilizer measurement in the second round of QEC yields $-1$, in which case the stabilizers need to be measured again; this adds another eight two-atom gates and four three-atom gates to the total resource cost.

\section{Computing Rydberg Blockade Radius Requirements for Rydberg FTQC Protocols}
\label{sec:blockade-app}

To obtain the blockade radius requirement for the Rydberg FTQC protocols, we must identify each physical qubit with an atom on the lattice, and then determine the maximum distance between two atoms which must interact with each other during a Rydberg gate. When the underlying atoms are placed in a triangular lattice, Figure~\ref{fig:logical-qubit-numbered} depicts convenient identifications for both the Ryd-7 and Ryd-3 codes. In this figure, numbers are used indicate the indices of data atoms within each logical qubit. (The index of a physical qubit within each logical qubit is the position, counting from the left, of that qubit in the definition of the logical states; see Equations~(\ref{eq:steane-0}) and (\ref{eq:steane-1}) for the seven-qubit code, or Equation~(\ref{eq:repetition-encoding}) for the three-qubit code.) 

In the Ryd-7 protocol, the blockade radius is defined by the interaction range needed to perform a logical $\CCZ$ gate between three neighboring logical qubits such as $A$, $B$, and $C$. Using the specific protocol given in Algorithm~\ref{alg:7q-ccz}, which involves 27 physical $\CCZ$ gates between atoms $j_A, k_B, l_C \in \{1,2,3\}$, we find the largest interaction range is required to perform the physical $\CCZ$ gate between farthest-separated triples such as $(j_A, k_B, l_C) = (3, 3, 1)$. For this specific case, the distances between atom pairs are $\textrm{dist}(j_A,k_B) = 3d$, $\textrm{dist}(j_A,l_C) = \sqrt{(7/2)^2+3/4}d \approx 3.61d$, and $\textrm{dist}(k_B,l_C) = 4d$. To apply the three-qubit Rydberg gate $R(j_A,k_B;l_C)$ as defined in Section~\ref{sec:summary}, this would require a blockade radius of $R_B > 4d$. However, this is not entirely necessary for our purposes: instead, it is sufficient that {\it two} out of the three distances $\textrm{dist}(j_A,k_B)$, $\textrm{dist}(j_A,l_C)$, and $\textrm{dist}(k_B,l_C)$ be less than the blockade radius. To see this, let us suppose, for example, that the distance between the two control atoms $j_A$ and $k_B$ is greater than $R_B$. In this case, applying the same pulse sequence as illustrated in Figure~\ref{fig:collective-gates}b would result in a three-qubit gate $R = \textrm{diag}(1,-1,-1,-1,-1,-1,1,1)$, which can also be obtained from the $\CCZ$ gate by single-qubit unitaries ($R \propto Y_1 Y_2 (\CCZ) X_1 X_2$).

The argument above allows the blockade radius requirement for Ryd-7 to be reduced to $R_B > 3.61d$ (thinner, light grey dotted line in Figure~\ref{fig:logical-qubit-numbered}). In fact, by modifying the implementation of the logical $\CCZ$ gate, it is possible to further reduce this requirement to $R_B > 3d$ (thicker, dark grey dotted line in Figure~\ref{fig:logical-qubit-numbered}); this is shown in Appendix~\ref{sec:ryd7-blockade-app}.

In the Ryd-3 protocol, the blockade radius $R_{B,1}$ is determined by the interaction range required to implement the logical Toffoli gate between neighboring logical qubits (e.g.,\ $A$, $B$, and $C$ in Figure~\ref{fig:logical-qubit-numbered}). As discussed in Appendix~\ref{sec:resource-cost-app}, there are two possibilities in this case. To directly implement every physical bias-preserving Toffoli gate using the circuit of Figure~\ref{fig:bias-preserving-toffoli}, the distance between $2_B$ and $A_3$ must be less than $R_{B,1}$; this requires $R_{B,1} > \sqrt{(7/2)^2+3/4}d \approx 3.61d$ (thinner, light grey dotted line in Figure~\ref{fig:logical-qubit-numbered}). However, this requirement can be reduced to $R_{B,1}>3d$ (thicker, dark grey dotted line in Figure~\ref{fig:logical-qubit-numbered}) at the expense of four additional two-atom entangling gates per logical Toffoli or Hadamard operation.

\section{Blockade Radius Reduction for Ryd-7}
\label{sec:ryd7-blockade-app}

To reduce the blockade radius requirement from $R_B = 3.61d$ to $R_B = 3d$ in the Ryd-7 protocol, we must modify the implementation of the logical $\CCZ$ operation. Recall that  Algorithm~\ref{alg:7q-ccz} implements a logical $\CCZ$ gate using 27 physical $\CCZ$ gates between the first three physical qubits of every logical qubit. This round-robin decomposition makes use of Eq.~(\ref{eq:ccz-round-robin}), which we now derive:
\begin{equation}
\label{eq:ccz-app}
\CCZ_{ABC} = \prod_{j_A,k_B,l_C \in \{ 1,2,3\}} \CCZ(j_A, k_B, l_C),
\end{equation}
To begin the derivation, we first recall that the logical states (\ref{eq:steane-0}) and (\ref{eq:steane-1}) of the seven-qubit code have well-defined parity: the number of physical qubits in the $|1\rangle$ state is always even for $|0\rangle_L$ and odd for $|1\rangle_L$. It then follows that the logical $\CCZ$ gate can be implemented in a fully round-robin fashion involving all physical qubits
\begin{equation}
\label{eq:rr-1}
\CCZ_{ABC} = \prod_{j_A,k_B,l_C \in \{ 1,2,...,7\}} \CCZ(j_A, k_B, l_C).
\end{equation}
This is because the round-robin implementation results in a $-1$ phase accumulation for each triple $(j_A,k_B,l_C)$ of physical qubits in the $|1\rangle$ state, and the number of such triples is odd if all logical qubits are in the $|1\rangle_L$ logical state, while it is even if at least one logical qubit is in the $|0\rangle_L$ state. To reduce this to Eq.~(\ref{eq:ccz-app}), we notice that for each choice of $j_A$ and $k_B$, the product
\begin{equation}
\label{eq:ccz-stab-1}
\prod_{l_C \in \{4,5,6,7\}} \CCZ(j_A, k_B, l_C)
\end{equation}
acts as an identity operation on the logical qubits, because $g_4 = Z_4 Z_5 Z_6 Z_7$ is a stabilizer of the seven-qubit code. We then multiply both sides of Eq.~(\ref{eq:rr-1}) by this operator, and use the fact that all the $\CCZ$ gates commute with each other and square to the identity operator. In this way, the product over $l_C$ in the logical $\CCZ$ gate can be reduced from $l_C \in \{ 1,2,...,7\}$ to $l_C \in \{ 1,2,3\}$. Because the $\CCZ$ gate is symmetric in the three involved qubits, this same argument can be applied to reduce the products over $j_A$ and $k_B$ to obtain Eq.~(\ref{eq:ccz-app}). 

To reduce the blockade radius requirement from $R_B = 3.61d$ to $R_B = 3d$, we can replace the product (\ref{eq:ccz-stab-1}) by
\begin{equation}
\label{eq:ccz-stab-2}
\prod_{l_C \in \{1,2,4,7\}} \CCZ(j_A, k_B, l_C)
\end{equation}
in our derivation for one of the logical qubits, say qubit $C$. This is because the single-qubit operator $Z_1 Z_2 Z_4 Z_7 = g_2 g_3$ is the product of two stabilizers, so the operator (\ref{eq:ccz-stab-2}) also acts trivially on the logical subspace. It follows that
\begin{equation}
\label{eq:rr-2}
\CCZ_{ABC} = 
\prod_{\substack{j_A,k_B \in \{1,2,3\} \\ l_C \in \{3,5,6\}}} \CCZ(j_A, k_B, l_C).
\end{equation}
Thus, the 27 physical $\CCZ$ gates in Algorithm~\ref{alg:7q-ccz} may be replaced by the 27 $\CCZ$ gates used in the right hand side of Eq.~(\ref{eq:rr-2}).

Given the geometrical layout of individual atoms within each logical qubit shown in Figure~\ref{fig:logical-qubit-numbered}a, we see that the required interaction range for implementing the logical $\CCZ$ operation using these 27 gates is smaller than the interaction range required to perform the 27 gates of Algorithm~\ref{alg:7q-ccz}. Furthermore, following the observation made in Appendix~\ref{sec:blockade-app}, we notice that these $9$ physical qubits need not all be within the blockade radius of each other, so long as every physical qubit $j_A \in \{1,2,3\}$ is within distance $R_B$ of every $l_C \in \{3,5,6\}$, and every $k_B \in \{1,2,3\}$ is within distance $R_B$ of every $l_C \in \{3,5,6\}$. This requirement is satisfied for any $R_B > 3d$, as shown in Figure~\ref{fig:logical-qubit-numbered}a.

\section{Square Lattice Geometry for Ryd-3}
\label{sec:ryd3-sq-app}

\begin{figure}[!t]
\includegraphics[width=0.75\textwidth]{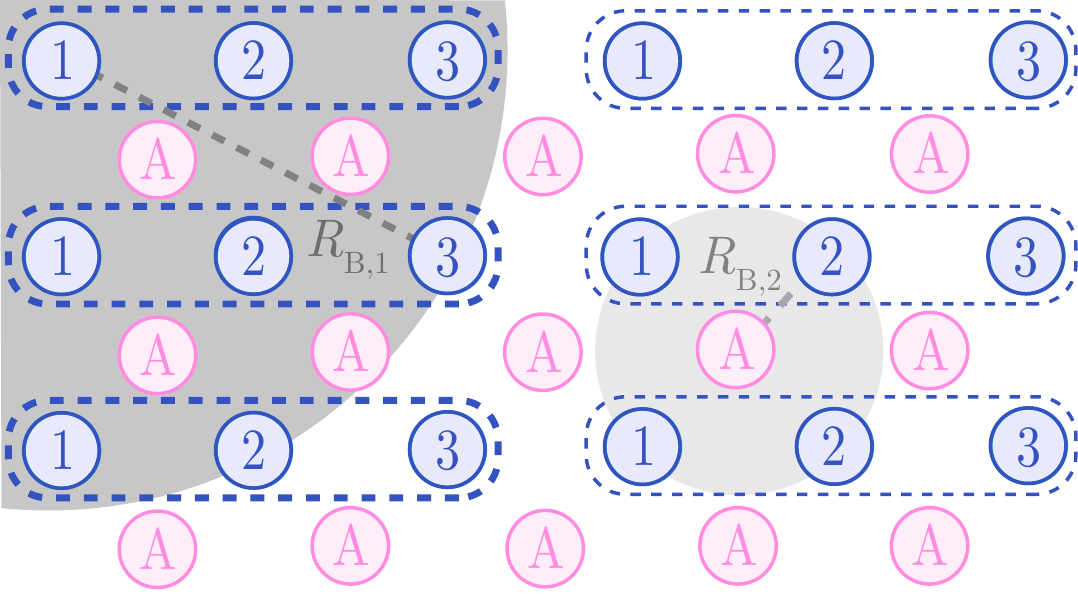}
\caption{Square lattice geometry for the Ryd-3 FTQC protocol. Data (blue) and ancilla (pink) atoms are placed on the vertices of a square lattice 
in an alternating fashion, with three data atoms comprising a logical qubit (blue dotted boxes). The numbers on each data atom indicate the index of that atom within each logical qubit; this is relevant for the implementation of stabilizer measurements and logical operations. Two Rydberg states with different blockade radii are required to implement the bias-preserving CNOT and Toffoli gates. The larger blockade radius $R_{B,1}$ must be larger than $\sqrt{10} d$ (dark grey), where $d$ is the nearest-neighbor spacing on the lattice, while the smaller blockade radius must satisfy $d < R_{B,2} < 2d$ (light grey). The interaction range $R_{B,1}$ is needed to perform a logical CCZ gate between the three logical qubits indicated in bold.}
\label{fig:three-qubit-sq}
\end{figure}

As mentioned in Section~\ref{sec:three-atom-geometry} of the main text, the Ryd-3 protocol can also be implemented when the underlying physical atoms are placed on a square lattice. In this case, the data and ancilla atoms are placed on the vertices of the lattice in an alternating fashion as shown in Figure \ref{fig:three-qubit-sq}. The stabilizer measurements can be implemented as discussed in Section~\ref{sec:ftqc-three-atom} if the smaller blockade radius $R_{B,2}$ satisfies $d < R_{B,2} < 2d$. The logical operation requiring the largest interaction range is the logical $\CCZ$ gate
\begin{equation}
\CCZ_{ABC} = \prod_{j_A,k_B,l_C \in \{ 1,2,3\}} \CCZ(j_A, k_B, l_C),
\end{equation}
which is implemented from 27 physical $\CCZ$ gates. To implement each physical gate, the distance between every pair $(j_A, l_C)$ and $(k_B, l_C)$ must be less than the larger blockade radius $R_{B,1}$, as discussed in Appendix~\ref{sec:blockade-app}. The longest such distance is $\sqrt{10}d$ as shown in the dark grey dotted line of Figure~\ref{fig:three-qubit-sq}, so the corresponding blockade radius requirement for this geometry is $R_{B,1} > \sqrt{10} d$.

With these blockade radii, the protocols of Section~\ref{sec:ftqc-three-atom} can be directly applied to perform all logical operations. We note that the higher density of ancilla atoms in this arrangement allows us to implement every physical Toffoli gate in the logical Toffoli operation directly using the circuit of Figure~\ref{fig:bias-preserving-toffoli}, without the need for additional ancilla atoms or CNOT gates (as was the case for two physical Toffoli operations under the triangular lattice geometry). In this way, for the square lattice geometry, the number of two-qubit entangling operations required for the logical Hadamard or Toffoli operations may be reduced by 4 compared to the numbers shown in Table~\ref{tab:gate-comparison}.

\section{Optical Pumping Procedure for the Bias-Preserving CNOT}
\label{sec:optical-pumping-app}

To implement the bias-preserving CNOT pulse sequence shown in Figure~\ref{fig:bias-preserving-cnot-target} of the main text, it is important that the optical pumping procedure in the final step pumps only the $m_F > 0$ states to the $|1\rangle$ state, and only the $m_F<0$ states to the $|0\rangle$ state. This requirement is essential to ensuring that the CNOT does not generate any $X$- or $Y$-type errors. For magnetic field regimes typically used in alkali atom Rydberg experiments, this state selectivity may not be straightforward to implement, as the level separation between different $m_F$ states within a single hyperfine manifold may be much smaller than the linewidth of the lasers used for optical pumping. To address this challenge, we can utilize a Rydberg state as a shelving state (due to its long lifetime) to avoid unwanted pumping of $m_F < 0$ (respectively, $m_F > 0$) states to $|1 \rangle$ ($|0 \rangle$). Thus, in Step 6 of Figure~\ref{fig:bias-preserving-cnot-target}, the optical pumping of $m_F > 0$ states into the $|1\rangle$ state can be implemented for $^{85}$Rb as follows:
\begin{enumerate}
	\item Swap the population between the $|1 \rangle$ state and the stretched ground state $|F=I+1/2, m_F=I+1/2 \rangle$.
	\item Swap the population between the $|0 \rangle$ state and the ground state $|F=3,m_F=0 \rangle$.
	\item Apply a resonant $\pi$ pulse to shelve any population in the $|F=2,m_F=-2 \rangle$ state into the Rydberg state $|nS_{1/2},m_J=-1/2,m_I=-5/2\rangle$.
	\item Use $\sigma^+$ light to excite states in the $F=2$ ground state manifold to the $5P_{3/2}$ $F=3$ manifold; these excited states decay quickly back to the ground state.
	\item Apply resonant $\pi$ pulses $|F=3, m_F=1 \rangle \leftrightarrow |F=2,m_F=1 \rangle$ and $|F=3,m_F=2 \rangle \leftrightarrow |F=2,m_F=2 \rangle$.
	\item Repeat Steps 4 and 5 as necessary; after several iterations, all population that started with $m_F > 0$ will be in the $|F=3,m_F=3 \rangle$ state.
	\item Repeat Steps 1, 2, and 3.
\end{enumerate}

Because the $|F=2,m_F=-2 \rangle$ state can only be populated if a Rydberg error occurred in one of the earlier steps of the bias-preserving CNOT, to leading order in the total error probability, we may assume that the Rydberg state $|nS_{1/2},m_J=-1/2,m_I=-5/2\rangle$ will not decay if it is populated in the above procedure. In this way, the only $F=2$ states that can be populated at the beginning of Step 4 above will be the $m_F > 0$ states, so the optical pumping will work in the same way as the protocol described in Section~\ref{sec:ftqc-lf} (Figure~\ref{fig:optical-pumping}). 
An analogous procedure can then be applied to pump the $m_F <0$ states into $|0 \rangle$. In this latter case, it will not be necessary to shelve population in the Rydberg state, as all $m_F > 0$ population will already have been transferred to the $|1 \rangle$ state.

\end{appendix}

\nocite{apsrev41Control}
\bibliographystyle{apsrev4-1}
\bibliography{refs}

\end{document}